\documentclass[sigconf]{acmart}

\usepackage{fontawesome}
\usepackage{tabularx}
\usepackage{multirow}
\usepackage{lscape}
\usepackage[capitalize, noabbrev]{cleveref}

\AtBeginDocument{%
  \providecommand\BibTeX{{%
    \normalfont B\kern-0.5em{\scshape i\kern-0.25em b}\kern-0.8em\TeX}}}

\copyrightyear{2024} 
\acmYear{2024} 
\setcopyright{rightsretained} 
\acmConference[DIS '24]{Designing Interactive Systems Conference}{July 1--5, 2024}{IT University of Copenhagen, Denmark}
\acmBooktitle{Designing Interactive Systems Conference (DIS '24), July 1--5, 2024, IT University of Copenhagen, Denmark}\acmDOI{10.1145/3643834.3660681}
\acmISBN{979-8-4007-0583-0/24/07}

\begin{document}

\title{Collage is the New Writing: Exploring the Fragmentation of Text and User Interfaces in AI Tools}

\author{Daniel Buschek}
\email{daniel.buschek@uni-bayreuth.de}
\orcid{0000-0002-0013-715X}
\affiliation{%
  \institution{University of Bayreuth}
  \city{Bayreuth}
  \country{Germany}
  \postcode{95447}
}

\renewcommand{\shortauthors}{Buschek}

\renewcommand{\shorttitle}{Collage is the New Writing: Fragmentation of Text and UIs in AI Tools}

\newcommand\daniel[1]{\textit{\textcolor{orange}{[Daniel] #1}}}

\newcommand{\numexamples}{36}
\newcommand{\numpapers}{32}

\newcommand\revised[1]{#1}

\begin{abstract}
  This essay proposes and explores the concept of \textit{Collage} for the design of AI writing tools, transferred from avant-garde literature with four facets: 1) \textit{fragmenting text} in writing interfaces, 2) \textit{juxtaposing voices} (content vs command), 3) integrating \textit{material from multiple sources} (e.g. text suggestions), and 4) shifting from manual writing to \textit{editorial and compositional decision-making}, such as selecting and arranging snippets.  
\revised{The essay then employs \textit{Collage} as an analytical lens to analyse the user interface design of recent AI writing tools, and as a constructive lens to inspire new design directions. 
Finally, a critical perspective relates the concerns that writers historically expressed through literary collage to AI writing tools. 
In a broad view, this essay explores how literary concepts can help advance design theory around AI writing tools.} It encourages creators of future writing tools to engage not only with new technological possibilities, but also with past writing innovations.
\end{abstract}



\begin{teaserfigure}
  \includegraphics[width=\textwidth]{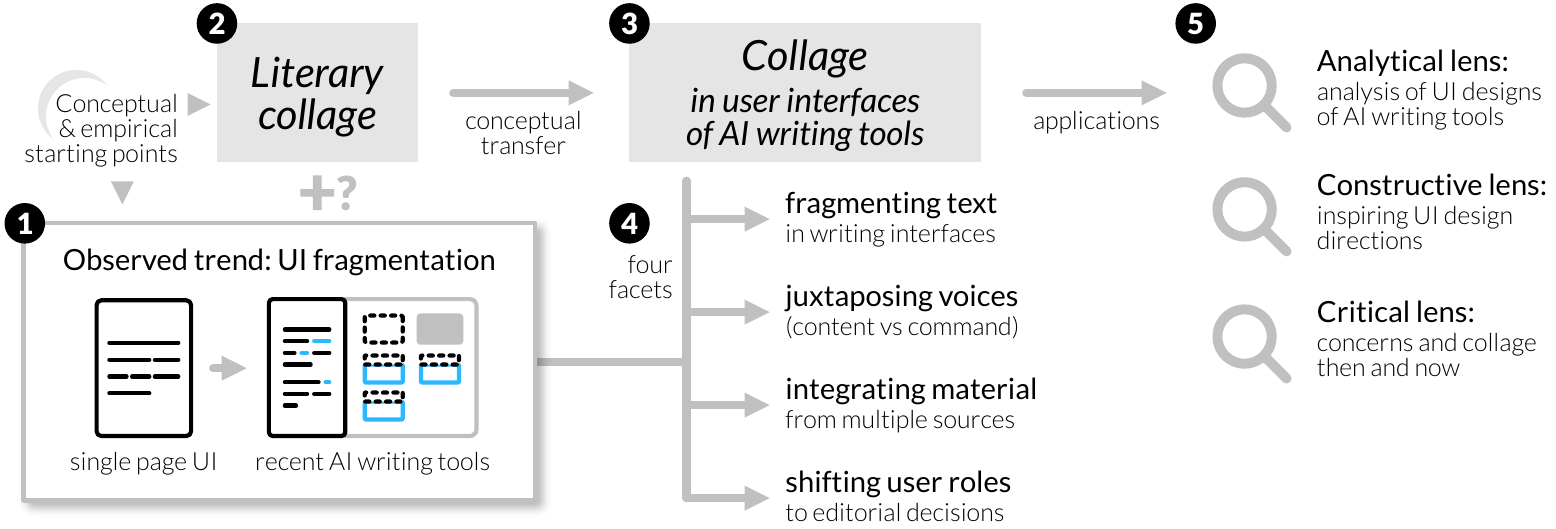}
  \caption{This essay \textbf{(1)} notices a trend towards fragmented text display in recent AI writing tools, compared to traditional UIs with a single page view. \textbf{(2)} Conceptually, text fragmentation also occurs in literature, such as in the \textit{literary collages} of avant-garde writing. \textbf{(3)} Investigating this connection, this essay transfers the concept of collage to the design of AI writing tools, \textbf{(4)} along four conceptual facets. \textbf{(5)} It then applies the resulting concept, referred to as \collage{}, in three ways -- as an analytical lens on existing designs, as a constructive lens to inspire new designs, and as a critical lens on broader concerns around writing with AI.}
  \Description{Visualised structure of this essay, from left to right: Illustration of a page UI view with lines to indicate text, plus an arrow leading to a second UI that looks a lot more complex, with many text boxes. This overall illustration, along with a text box saying ``Literary collage'' is connected with an arrow saying ``conceptual transfer'' to another box saying ``Collage in UIs of AI writing tools''. Below this box, arrows and text list the four identified facets of the transferred concept of Collage: fragmenting text, juxtaposition of voices, integrating material from multiple sources, and shifting user roles. Finally, another arrow saying ``applications'', points to three lens icons and corresponding text: analytical lens, constructive lens, critical lens.}
  \label{fig:teaser}
\end{teaserfigure}

\newcommand{\collage}{\textit{Collage}}

\maketitle

\section{Introduction}
Writing is being rewritten, as we write. There are many examples: When the novel was novel it did not conform to contemporary forms. It was the new way of writing, hence its name~\cite{shields2011reality}. 
In the 1920's, the avant-garde wrote ahead on radical new forms, breaking established structures.
Such new writing emerges from culture and art, and also technology, in a task-artifact cycle~\cite{carroll1991task}, from the printing press to the typewriter to digital word processing~\cite{kirschenbaum2016track}. 
Today, with the rise of generative AI, writing tools produce words that ask us: 
What is the new writing now? And how might we design tools for it?

The Human-Computer Interaction (HCI) community has reacted to the rise of Large Language Models (LLMs) with a surge of design instances. However, we currently lack abstracted design knowledge or \textit{strong concepts}, as described by \citet{hook2012strongconcepts}. These concepts complement particular design examples to develop a more generalised understanding of the involved HCI phenomena through ``discursive knowledge construction''. This essay proposes \collage{} as a contribution to such a discourse on intermediate-level concepts for the design of user interfaces (UIs) of AI writing tools.%

The key proposition is that literary collage, taken from the avant-garde repertoire, can be meaningfully transferred to the design of AI writing tools. 
\revised{The goal of this is to make explicit the design concepts and principles of fragmentation in both the \textit{UI design} of AI tools and in their implied \textit{writing processes}; more so than in final texts, which neither tool designers nor users typically desire to be perceived as fragmented today.}

In this essay, we refer to the transferred concept as \collage{} and to the original as \textit{literary collage}.
Concretely, we examine four conceptual facets that connect literary collage and HCI:

\begin{description}%
    \itemsep.5em
    \item[UI:] UIs for writing increasingly afford and demand \textbf{\textit{fragmenting text}} (e.g. multiple text entry areas), thus tearing apart the traditional page view that used to be the single focus of writing, such as in \textit{Microsoft Word}.
    \item[Interaction:] Interacting with AI writing tools requires users to \textbf{\textit{juxtapose different voices}}, as they switch between writing \textit{on} their text and writing \textit{about} it, in particular to the AI. \citet{dang2023choice} referred to these as \textit{diegetic} (e.g. draft/content text such as ``Once upon a time, Alice...'') and \textit{non-diegetic} (e.g. prompts/instructions, such as ``Write about Alice's adventures''). This essay uses these terms as well. %
    \item[Artefact:] The text may integrate \textbf{\textit{material from multiple sources}} -- also sources other than the author's own work. For example, parts of the text might be AI-generated, which sometimes might more or less closely resemble instances in the system's training data.
    \item[User:] Many AI tool designs shift the user's role from manual writing to \textbf{\textit{making editorial and compositional decisions}}, such as selecting and arranging generated text snippets.
\end{description}

\revised{In summary, both literary collage and AI writing tool UIs fragment text, juxtapose voices, integrate material from multiple sources, and shift the writer's role towards editorial decisions.}

With these concrete aspects (described in detail in \cref{sec:collage}), we employ the resulting concept of \collage{} in three ways: With the \textit{analytical} lens (\cref{sec:analytical_lens}), we analyse UI designs of recent AI writing tools. With the \textit{constructive} lens (\cref{sec:constructive_lens}), we explore design directions inspired by embracing or rejecting \collage. Finally, with a \textit{critical} lens (\cref{sec:critical_lens}), we reflect on issues around writing with AI by engaging with the concerns that writers historically expressed through literary collage. \cref{fig:teaser} shows this structure.

In a broad view, this essay explores \revised{how literary concepts can help advance design theory around AI writing tools}. It encourages creators of future writing tools to engage not only with new technological possibilities, but also with past writing innovations.
\section{Background \& Related Work}

This section clarifies the background and scope of this essay and relates it to work from HCI, Media Sciences, and AI, which engages with literature and writing.

\subsection{AI Tools for Writing}

This essay uses ``AI tools for writing'' or ``AI writing tools'' to broadly refer to systems that are interactive and designed for writing digital texts. Here, ``interactive'' means that human input of some kind is a key part of the design and intended application. This will often be iterative and in particular excludes fully automated text generation systems that do not involve users or a UI. Moreover, ``writing digital texts'' is intended to capture both 1) all sorts of texts (e.g. chat messages, emails, blog posts, business reports, novels, poems), and 2) all stages of writing (e.g. pre-writing, drafting, revising, editing). %

Related, we use ``AI tool'' in a broad sense, akin to ``interactive AI system''. The term ``tool'' in this essay is \textit{not} intended to indicate a specific delineation from HCI conceptualisations such as ``agent'' or ``asssistant'', as known from HCI debates~\cite{shneiderman1997debate, farooq2017debate}. In this sense here, the currently popular ChatGPT\footnote{\url{https://openai.com/chatgpt}} system is an AI writing tool although in many ways it is presented as an agent (e.g. through conversational interaction, ``writing'' text over time, using ``I'').

For a recent survey of writing tools/assistants, see the work by~\citet{lee2024dsiiwa}. This essay also covers a range of examples, most prominently in the analysis of \numexamples{} published designs in \cref{sec:analytical_lens} (see \cref{tab:paper_set}).

\subsection{Literary Collage \& Links to Writing with AI}\label{sec:related_work_lit_collage}

\citet{vietta1974} defines literary collage in the context of avant-garde literature, specifically Expressionism, as ``units assembled from other, `prefabricated' texts. The author no longer produces them personally, but simply assembles [them]''.
Related, the glossary of the Poetry Foundation~\cite{poetryFoundation2024collage} states that ``collage in language-based work can now mean any composition that includes words, phrases, or sections of outside source material in juxtaposition.''
Besides using such external text material, collage may also juxtapose different writing styles or voices, even if written by one author~\cite{shields2011reality}. In both these ways of defining literary collage, it refers to a fragmentary character of the work. \revised{The conceptual aspects in these definitions guided the identification of the four facets here (\cref{sec:collage}).}

Literary collage can be a form (e.g. a collage novel) as well as a style or technique (e.g. using collage to achieve a fragmented impression). For this essay, this distinction is not important and we refer to collage both as a form and a stylistic device.

From the above definitions, we can already spot connections to writing with AI tools: First, an LLM provides the ``prefabricated'' text pieces to be arranged by the user. Second, the link to ``outside source material'' acknowledges both the AI's training texts, as well as the fact that AI-provided text in the UI originated from ``outside of the user''. Third, throughout interacting with this system, the user might have to switch frequently between entering text in different voices. The two key voices are: writing on the draft (e.g. ``Once upon a time''), and writing instructions to the AI (e.g. ``Make this paragraph more engaging''). \citet{dang2023choice} referred to these as \textit{diegetic} and \textit{non-diegetic} text entry in interaction with LLMs.

Finally, there is a key difference in scope between literary collage and our transferred concept of \collage{}: Literary collage produces \textit{perceivably} fragmented texts. In contrast, users of AI writing tools may write many things that do not look fragmented (e.g. typical emails, blog posts, business reports, novels, etc.). %
This essay focuses not on the fragmentation of a final text, but on the fragmentation of text elements in UIs and interaction processes involved in writing with AI. Nevertheless, some aspects discussed here also relate to the final output (e.g. parts of text entered by the user, others generated by AI), although this might not necessarily be visible to the readers. \revised{In that view, \collage{} can encourage revealing such ``seams'' (\cref{sec:design_exploration1}), also in future UIs for document viewers. For instance, these might reveal mixed (human/AI) authorship on a more fine-grained level than a document-wide statement (cf.~\cite{draxler2023ghostwriter}).} %

\subsection{At the Intersection of Literature, HCI \& AI}

In general, new writing tools have had a huge impact on writers throughout history. The book by \citet{kirschenbaum2016track} tracks these changes over time, discussing how (famous) writers reacted to the development of word processing devices and software.

\subsubsection{Avant-garde and Interactive Software}
The idea of connecting artistic/literary forms with a reflection on technological developments is not new: Most closely related to this essay, \citet{manovich1999avantgarde} described ``avant-garde as software'' in 1999. Among several parallels, he noted how ``the avant-garde strategy of collage re-emerged as a `cut and paste' command''. Other examples include the simultaneous interaction with multiple windows, which the user can move, also into overlapping arrangements. Manovich relates this mostly to cinematic montage, both temporal and within a shot. This essay also picks up on avant-garde strategies in interactive software but explores them not for visual art and any application but more specifically for writing and software tools that use generative AI.

\subsubsection{Enabling Text Assembly by Separating Data from Structure}
Regarding text, \citet{manovich1999avantgarde} also highlighted how hyperlinking separates data from structure -- the same (linked) pieces can be integrated in various places to assemble a larger message. In this light, LLMs could be seen as a catalyst for an extreme version of such ``message construction'': Manovich's pieces were ``atomistic (i.e. discrete)'', such as linking to a whole section or website. In contrast, today's LLMs separate input data from its structure much more dynamically. Trained to predict the next words from preceding (con)text, they can provide text pieces that are dynamically shaped to fit into the user's draft in terms of grammar, style, tone, and so on. Currently, this fluidity often comes at the expense of tracing back sources, in stark contrast to hyperlinks. 
We return to this in \cref{sec:constructive_lens}, where we explore how embracing the seamful nature of \collage{} could inspire new UI designs. %

\subsubsection{Fiction and Design in HCI}
In HCI and creativity research, \citet{kantosalo2021cent18th} explored 18th century literature through a lens of design fiction. Specifically, they investigated embodiment of human-machine co-creativity in five selected texts. One example is Hoffmann's ``Automata'' (1814), in which a mechanical ``Talking Turk'' answers questions.\footnote{Another link to HCI: A (fake) automata built at the time is the chess player by Wolfgang von Kempelen, which inspired the name of Amazon's \textit{Mechanical Turk} system~\cite{schwartz_untold_2019}, used in many HCI studies over the last years.} 
Their paper concludes with ``four guidelines for discovery'' and their methodological reflection emphasises the role of ``historical fiction as sources of inspiration for future [...] research''. There are also calls for broader engagement with historical contexts in and around HCI~\cite{bodker2023hcihistory, forlano2023speculativehistories}. This essay also revisits historical writing to inspire future research. In contrast to the related work, the focus is on a form or style (literary collage, fragmentation) and not on specific fictional works or a broad historic picture. %
Nevertheless, \cref{sec:critical_lens} draws connections from historical concerns expressed through literary collage. 
Finally, with a focus on interaction, this essay is also loosely related to the work by \citet{axtell2021startrek}, who investigated speech interaction design as envisioned in works of science fiction (\textit{Star Trek}).%

\subsubsection{Fiction as a Lens on LLMs}
AI researchers have also used literature as a lens to reflect on LLMs. Concretely, \citet{bottou2023borges} have recently likened the space of possible texts, represented by an LLM, to themes in the work of Jorge Luis Borges, specifically \textit{The Garden of Forking Paths} and \textit{The Library of Babel}. Each word entered or generated provides a potential fork -- a sentence could continue in a myriad of ways, probabilistically modelled by an LLM. The interaction design to access this space is simple -- type a few words. However, in the continued forking paths then generated by the LLM, ``nothing tells the true from the false, the helpful from the misleading, the right from the wrong''. That said, the LLM is not deceitful as it lacks intention. Instead, it ``merely follows the narrative demands of the evolving story''.
This motivates this essay on a fundamental level: If LLMs lend writing tools narrative momentum along literary forms, the design of their practical controls (UI) may benefit from a deeper engagement with literature, too.

In addition, \citet{bottou2023borges} observe that LLMs trained on the internet have processed science fiction. Thus, in interaction with a chatbot like ChatGPT, ``when our part of the dialog suggests that we wonder whether the machine is sentient, the machine's answers draw on the abundant science fiction material found in its training set.'' 
In other words, literature already directly impacts interaction and user experience with widely used LLM-based products today. 
This further motivates exploring connections to literature in detail; in our case here not as a part of model training, but for UI and interaction design.

\section{The Concept of Collage}\label{sec:collage}
This section describes the proposed concept of \collage{} in detail, as transferred from literary collage to the context of AI writing tools. \revised{Concretely, we examine four facets. These are the four specific, original constructs from literary collage (see~\cref{sec:related_work_lit_collage}) that we are applying in the new context of AI writing tools.} Here we relate them to the UI (fragmenting text), interaction (juxtaposing voices), the artefact/text (multiple sources), and the user (shifting roles towards editorial and compositional decision-making).

\subsection{UI: Fragmenting Text in Writing Interfaces}\label{sec:collage_fragmentation}

In UIs of traditional writing software (e.g. \textit{Microsoft Word}), the main view is a page, a metaphor of a physical sheet of paper~\cite{kirschenbaum2016track}. %
Other text on the screen might be visible in parallel, such as the labels in a toolbar, status bar or menu. However, this UI text is not specifically connected to the draft text. %
Some tools also offer annotations, such as margin comments for collaborative writing. These afford text entry that might be relevant for the work on the main text (e.g. entering a suggested revision for your co-authors). %
That said, even in these cases, the page view typically covers most of the screen space. In short, the page has been a largely uncontested UI centre-piece where writing happens. 

This is no longer the case for many recent writing tools built on generative AI. Instead of a single locus of writing, their UIs are composed of \textit{fragmented} text views. These are given more equally distributed importance in terms of their screen space and impact on the writing outcome (e.g. see~\cite{hoque2024hallmark, huang2023inspo, goodman2022lampost}). This fragmentation is the first of the four facets of \collage{}, as proposed here.

UI metaphors are shifting accordingly, from one sheet of paper to a multitude of cards, boxes, and snippets (e.g.~\cite{dang2022beyond, hoque2024hallmark, huang2023inspo, kim2023cellsgenlens, zhang2023visar}). The page has been torn apart. As UI elements, the pieces can now be added, moved, removed, combined, split, and so on. Thus, such UIs afford and demand \collage{} -- by designers, users, or both.

Concretely, in some UIs, the pieces are arranged by the UI designers and thus largely fixed during use (e.g.~\cite{biermann2022tooltocompanion, mirowski2023drama, singh2023elephant}). %
In other cases, users can create and work with them dynamically (e.g.~\cite{dang2022beyond, kim2023metaphorian, lin2024rambler, zhang2023visar}).%

\textbf{In summary: Regarding UI design,} \collage{} highlights both the fragmented design of writing surfaces in UIs, as well as their affordance of the users' acts of fragmenting text (and de-fragmenting it, i.e. assembling). %
As such, this is a descriptive insight -- fragmentation might be good or bad for users. However, the perspective of \collage{} surfaces fragmentation as a relevant cross-cutting aspect of UI design and gives it a name. %
As laid out above (and in more detail in \cref{sec:analytical_lens}), fragmentation shows up in many recent AI writing tools. If we recognise this as an emerging key affordance and/or user need, it is worth making it explicit in design. %

\subsection{Interaction: Juxtaposing Voices in Writing}\label{sec:collage_voices}

In traditional UIs for writing, there is one main type or ``voice'' of text entry: writing on the draft. %
This is no longer the case for writing with generative AI tools that support prompting (i.e. writing commands/instructions to the AI)~\cite{faltings2021commandediting}. With these, users switch between writing in two voices: writing on the draft vs writing prompts (i.e. diegetic vs non-diegetic writing~\cite{dang2023choice}).

Recent tool designs frequently promote using both in a tight interaction loop, such as via parallel UI views. For example, \textit{Wordcraft} combines a page view with prompting in a sidebar ``to engage in open-ended conversation about the story [with] custom requests expressed in natural language''~\cite{yuan2022wordcraft}. Other examples include \textit{LaMPost}~\cite{goodman2022lampost} and \textit{HaLLMark}~\cite{hoque2024hallmark}.
\citet{dang2023choice} integrate both voices even more closely: Without using the mouse, users can switch from drafting to entering prompts that refine AI sentence suggestions.

Designs like these ask writers to enter text in both voices in \textit{combination} and \textit{iteratively}, to fully benefit from the AI features. This in turn implies switching between voices \textit{frequently}. The resulting juxtaposition of writing in different voices is the second of the four facets of \collage{} as proposed in this essay.

To examine this more closely, we note that even without generative AI, writers need to switch between text entry for their draft and other actions, such as using a formatting toolbar. There is a key difference, though: For the toolbar, the user needs to recognise an icon and click on it. In contrast, when using a prompting-based AI feature, the user needs to think about how to compose an effective prompt. This is challenging because writing and prompting require overlapping cognitive processes, such as coming up with an idea and translating it into a series of words, before entering those physically~\cite{flower1981writingprocess, hayes2012modelingwriting}. Thus, drafting and delegating in this way are hardly parallelisable, as supported by empirical and conceptual work~\cite{bhat2023suggestions, dang2022beyond, tankelevitch2023metacognitive}. This is also practically experienced by those who have felt distracted from their own writing by text suggestions in the UI, or who have found it more difficult to express their intention in a prompt than writing themselves (cf.~\cite{zamfirescu2023johnny}). %

Even users of traditional writing software might switch to parallel text-based tasks, such as searching in a browser. In that sense, switching  voices is not entirely new with AI. However, a browser has a separate application window. Similarly, a user might add comments to a draft for collaborators. Nevertheless, that is typically an optional feature, not an integral part of writing. %

\textbf{In summary: Regarding the interaction design,} \collage{} highlights the \textit{essential and close integration} of entering text in different voices (i.e. juxtaposition) that is facilitated and required by the tools' UI and interaction design. This comes with new opportunities and challenges: Users can now dynamically elicit desired writing support functionality via prompting. On the flip side, switching frequently can be difficult. Therefore, in the perspective of \collage{}, relieving this tension between dynamic appropriation and multitasking emerges as a central goal for successful interaction design of AI writing tools. %

\subsection{Artefact: Integrating Material from Multiple Sources}\label{sec:collage_sources}

Traditional writing software predominantly assumes that ``writing'' equals entering text on a page view. While copy+paste and import might support bringing in outside text pieces, these are not the expected main ``writing acts'' performed by the user. This is evident, for example, from the fact that few text editors provide a graphical representation of clipboard content, at least not as a main view.

In literary collage, the author assembles text from pieces taken from other sources, including writing of other people. Transferred to writing with AI, one may consider such other sources to be ``the AI'' itself, as well as the myriad texts used to train an LLM. %

One way for these sources to influence the user's writing are text suggestions~\cite{jakesch2023opinionated}. UIs with this feature typically show a sentence continuation directly on the page at the caret, or they list multiple suggestions in a popup window or sidebar (e.g.~\cite{buschek2021phrasesuggestions, dang2023choice, lee2022coauthor}). In this kind of influence, we include both the more implicit impact of the training texts on what is generated (and what the user makes of it when seen as a suggestion), as well as direct copying of training data in generated LLM output (e.g. see analyses~\cite{lee2023llmplagiarize} and lawsuits around this~\cite{grynbaum2024nytlawsuit}). \revised{In this light, some AI tools might introduce plagiarism, possibly unnoticed by users, which tool creators should aim to avoid, both on the modelling level, as well as through the UI and interaction design.}

A second avenue of influence of outside material is present in AI tools that explicitly support research into external sources. Such tools might find references (e.g. looking up similar published sentences~\cite{shen2023convxai, shen2023convxai2}). More subtly, this is also present in any UI that offers a free text prompt box: Even if the tool is not presented as an information retrieval tool, the user could use the prompt to request world information (e.g. ``What is the capital of France?'') -- a correct answer has to come from somewhere in the training data.

These possibilities for influencing the text through outside material from multiple sources, and their more or less direct integration, constitute the third of the four facets of \collage{}. 

Currently, the fact that many sources contribute to the model's generation does not typically show up in the UI and resulting text. Thus, it might be seen as not directly affecting UI and interaction design of such tools. 
In this light, \collage{} reminds us that the sources \textit{could} be shown. Indeed, some recent designs include visual indicators for this. For example, \textit{HaLLMark}~\cite{hoque2024hallmark} uses coloured highlighting for users to keep track of which text pieces were entered by them, and which ones were inspired or added by an LLM.
\revised{This is an example of fragmentation in line with \collage{}, which might help users avoid accidental plagiarism when writing with AI.}

\revised{More generally, this facet of \collage{} is \textit{not} intended to endorse plagiarism or misrepresentation of authenticity. On the contrary, by explicitly acknowledging the influence of external material, it aims to raise awareness of risks around plagiarism and accountability in such tools~\cite{Li2024valbencon}.}

\textbf{In summary: Regarding the artefact} (i.e. text) resulting from interaction with AI writing tools, \collage{} acknowledges the more or less visible influence of other sources of text material, that is, sources beyond the user/writer. This awareness could fruitfully inform UI and interaction design. %

\subsection{User: Shifting Role towards Editorial and Compositional Decision-making}\label{sec:collage_roles}

In traditional writing software, the user is in the role of the author. There is no draft if the user does not enter it.
Features that process this draft serve a supporting role, %
such as by checking spelling and grammar, listing synonyms, and detecting repeated words~\cite{burstein-wolska2003wordrep}.

On the one hand, AI such as LLMs can be used to improve these features (e.g. a thesaurus based on word embeddings~\cite{gero2019thesaurus}). To the extent that these evolved features do not substantially contribute or revise text, they seem unlikely to shift the roles of user and system. Users are still in the author's seat. A similar role distribution might be expected with features that evaluate text, yet do not change it, perhaps akin to comments from a reviewer (cf.~\cite{benharrak2023writerdefined}).

On the other hand, LLMs can power tools that generate or revise text, or otherwise take more initiative and/or agency over the evolution of a draft. In these cases, we might expect a (partial) shift in the user's perceived role. 

Indeed, this shift has implications for the user's experience of the interaction, as well as the resulting text. For instance, one of the interaction designs by \citet{lehmann2022sugglistcont} emphasised an editorial user role, which resulted in lower ratings on users' perceived authorship. Related, \citet{draxler2023ghostwriter} found an \textit{AI ghostwriter effect}: ``Users do not consider themselves the owners and authors of AI-generated text but refrain from publicly declaring AI authorship.'' This effect was mediated by the role of the user as a result of interaction design, as they found that ``subjective control over the interaction and the content increases the sense of ownership.''

This is in line with literary collage: Authors engaging in collage necessarily take on a more editorial role. Instead of writing, or in addition to it, they make decisions about the selection and composition of elements. These decisions can turn the result into a creative piece by that person, seen as more than the sum of its parts (cf.~\cite{manovich2002language, shields2011reality}). 

\textbf{In summary: Regarding the user,} \collage{} highlights the new roles that need to be supported through UI and interaction design in AI writing tools. These roles demand decision-making on a higher level of abstraction. Concretely, this involves work such as selecting material, making developmental editing choices, and (re-)arranging elements. In addition, this facet of \collage{} reminds us of the complexities of authorship that arise from the combination of these shifted roles and creative actions. In software tools, these are enabled and enacted by the UI and interaction design.

\subsection{Summary and Definition}
\revised{The described conceptual transfer amounts to this definition:
\collage{} in AI writing tools refers to \textit{fragmented} UI designs and their support for -- and/or demand of -- working with fragmented text pieces. This interaction \textit{juxtaposes writing in two voices}, namely diegetic text entry (i.e. writing on the draft) and non-diegetic text entry (i.e. writing prompts to the AI). UI and interaction involve \textit{material from multiple sources}, such as human vs AI-generated text, and by extension the AI's training data. Together, this \textit{shifts the user's role} from manual writing towards making editorial and compositional decisions.}

\section{Analytical Lens: Collage in AI Writing Tools}\label{sec:analytical_lens}

This section applies \collage{} as an analytical lens to analyse recent AI writing tools by considering the described four facets.

\begin{table*}[]
\centering
\scriptsize
\renewcommand{\arraystretch}{1.25}
\setlength\tabcolsep{4.75pt}
\newcolumntype{L}{>{\raggedright\arraybackslash}X}
\begin{tabularx}{\textwidth}{@{}Xcccccclclcclccc@{}}
\toprule
\multirow{3}{*}{\textbf{Paper}} & \multirow{3}{*}{\textbf{CF}} & \multicolumn{14}{c}{\textbf{Facets of \collage}} \\
 &  & \multicolumn{5}{c}{Fragmentation} &  & Multiple voices &  & \multicolumn{2}{c}{Multiple sources} &  & \multicolumn{3}{c}{Shifting roles} \\
 &  & In & Out & Spatial & Segment & Connect &  & Dieg. + non-dieg. &  & Provides text & Indicator &  & Explore & Arrange & Curate \\ 
 \midrule
    \citet{afrin2021student} & 1 & & & & \faCheck & &  & &  & & &  & & & \\ 
    \citet{arakawa2023catalyst} & 1 & & & & & &  & &  & \faCheck & &  & & & \\ 
    \citet{tsai2020lingglewrite} & 2 & & \faCheck & & \faCheck & &  & &  & & &  & & & \\ 
    \citet{bhat2023suggestions} & 3 & & & & \faCheck & &  & &  & \faCheck & &  & & & \faCheck \\ 
    \citet{buschek2021phrasesuggestions} & 3 & & \faCheck & & & &  & &  & \faCheck & &  & & & \faCheck \\ 
    \citet{gero2022sparks} & 3 & & \faCheck & & & &  & \faCheck &  & \faCheck & &  & & & \\ 
    \citet{hui2023lettersmith} & 3 & & & & \faCheck & &  & &  & \faCheck & &  & \faCheck & & \\ 
    \citet{jakesch2023opinionated} & 3 & & & & \faCheck & &  & &  & \faCheck & &  & & & \faCheck \\
    \citet{lee2022coauthor} & 3 & & \faCheck & & & & & &  & \faCheck & & & & & \faCheck \\
    \citet{lehmann2022sugglistcont} & 3 & & \faCheck & & & &  & &  & \faCheck & &  & & & \faCheck \\ 
    \citet{singh2023elephant} & 3 & & \faCheck & & \faCheck & &  & & & \faCheck &  & & & \\
    \citet{biermann2022tooltocompanion}, a & 4 & & \faCheck & & \faCheck & &  & &  & \faCheck &  &  & \faCheck & & \\
    \citet{biermann2022tooltocompanion}, e & 5 & \faCheck & \faCheck & & \faCheck & &  & &  & \faCheck & \faCheck &  & & & \\ 
    \citet{clark2018slogans} & 5 & \faCheck & \faCheck & & & &  & &  & \faCheck & &  & \faCheck & & \faCheck \\ 
    \citet{chung2022talebrush} & 5 & & & \faCheck & \faCheck & &  & &  & \faCheck & \faCheck &  & \faCheck & & \\  
    \citet{dang2022beyond} & 5 & & \faCheck & \faCheck & \faCheck & \faCheck &  & &  & & &  & & \faCheck & \\ 
    \citet{dang2023choice} & 5 & \faCheck & \faCheck & & & & & \faCheck &  & \faCheck & &  & & & \faCheck \\ 
    \citet{benharrak2023writerdefined} & 6 & \faCheck & \faCheck & & \faCheck & &  & \faCheck &  & \faCheck & \faCheck &  & & & \\ 
    \citet{biermann2022tooltocompanion}, b & 6 & \faCheck & \faCheck &  & \faCheck & &  & \faCheck &  & \faCheck & &  & \faCheck & & \\ 
    \citet{biermann2022tooltocompanion}, c & 6 & \faCheck & \faCheck & & \faCheck & &  & \faCheck &  & \faCheck & &  & & \faCheck & \\ 
    \citet{ito2023rewriting} & 6 & & \faCheck & & \faCheck & &  & &  & \faCheck & \faCheck &  & \faCheck & & \faCheck \\ 
    \citet{lin2024rambler} & 6 & \faCheck & & \faCheck & \faCheck & &  & \faCheck &  & \faCheck & &  & & \faCheck & \\ 
    \citet{peng2023storyfier} & 6 & \faCheck & \faCheck & & \faCheck & &  & &  & \faCheck & \faCheck &  & & & \faCheck \\ 
    \citet{biermann2022tooltocompanion}, d & 7 & \faCheck & \faCheck & \faCheck & \faCheck & \faCheck &  & &  & \faCheck & &  & & \faCheck & \\
    \citet{goodman2022lampost} & 7 & \faCheck & \faCheck & & & \faCheck &  & \faCheck &  & \faCheck & &  & \faCheck & & \faCheck \\
    \citet{kim2023metaphorian} & 7 & & \faCheck & \faCheck & \faCheck & \faCheck &  & &  & \faCheck & &  & \faCheck & & \faCheck \\  
    \citet{mirowski2023drama} & 7 & \faCheck & \faCheck & & \faCheck & &  & \faCheck &  & \faCheck & &  & \faCheck & & \faCheck \\ 
    \citet{shen2023convxai, shen2023convxai2} & 7 & \faCheck & \faCheck & & \faCheck & &  & \faCheck &  & \faCheck & \faCheck &  & \faCheck & & \\ 
    \citet{yuan2022wordcraft} & 7 & \faCheck & \faCheck & & & \faCheck &  & \faCheck &  & \faCheck & &  & \faCheck & & \faCheck \\ 
    \citet{huang2023inspo} & 8 & & \faCheck & & \faCheck & \faCheck &  & \faCheck &  & \faCheck & \faCheck &  & \faCheck & & \faCheck \\ 
    \citet{lu2019inkplanner} & 8 & & \faCheck & \faCheck & & \faCheck &  & &  & \faCheck & \faCheck &  & \faCheck & \faCheck & \faCheck \\
    \citet{park2023storytelling} & 8 & & \faCheck & \faCheck & \faCheck & \faCheck &  & &  & \faCheck & \faCheck &  & \faCheck & & \faCheck \\ 
    \citet{shi2023effidit} & 8 & \faCheck & \faCheck & & \faCheck & &  & \faCheck &  & \faCheck & \faCheck &  & \faCheck & & \faCheck \\ 
    \citet{hoque2024hallmark} & 9 & \faCheck & \faCheck & \faCheck & \faCheck & \faCheck &  & \faCheck &  & \faCheck & \faCheck &  & \faCheck & & \\ 
    \citet{kim2023cellsgenlens} & 9 & \faCheck & \faCheck & \faCheck & \faCheck & \faCheck &  & \faCheck &  & \faCheck & &  & \faCheck & & \faCheck \\ 
    \citet{zhang2023visar} & 10 & \faCheck & \faCheck & \faCheck & \faCheck & \faCheck &  & \faCheck &  & \faCheck & &  & \faCheck & \faCheck & \faCheck \\ 
 \bottomrule
\end{tabularx}
\caption{Overview of the \numexamples{} AI writing tools (from \numpapers{} papers) included in the analysis in \cref{sec:analytical_lens}. From left to right, the columns list the \textit{Paper} reference, the \textit{Collage Factor (CF)}, and eleven \textit{binary questions} across the four facets of \collage{} (\cref{sec:collage}). Each checkmark indicates a ``Yes'' for a tool and question. The CF is the sum of the checkmarks per paper (row). Tools/papers are ordered by CF, that is, by an increasing degree of \collage{} in their UI design. See \cref{sec:appendix_ui_analysis} for a detailed explanation of each question, \cref{sec:analysis_method} for the process, and \cref{sec:analysis_results} for the results.}
\Description{Table listing the reviewed papers (rows), with checkmarks for each question (columns) that a paper fulfills (i.e. ``yes'' answer). The overall impression is that most papers have a ``Yes'' for multiple AI outputs in general, visually segmenting text pieces, showing AI-generated text in the UI, and allowing the user to curate AI-generated text. About half the papers afford text entry for prompting the AI.}
\label{tab:paper_set}
\end{table*}

\subsection{Reviewing the Design of AI Writing Tools}\label{sec:analysis_method}

To \revised{ground this essay and} operationalise \collage{} as an analytical lens, this section links the facets to designs \revised{in the literature}. This is not a survey. Instead, it is intended to illustrate the analytical value of \collage{} \revised{and share more of the thinking behind it for concrete UI designs}.
With this goal, here we consider \numexamples{} examples from \numpapers{} papers that either contributed AI writing tools (e.g.~\cite{chung2022talebrush, dang2023choice, yuan2022wordcraft}) or built them to study their impact (e.g.~\cite{bhat2023suggestions, buschek2021phrasesuggestions, jakesch2023opinionated}).

The selection was motivated to cover 1) recent work, 2) widely referenced, impactful designs, 3) diverse tools (w.r.t. stages, styles, use cases), and 4) diverse UI designs (assessed from elements and layouts in screenshots). Papers were found via a recent survey of 115 writing assistants~\cite{lee2024dsiiwa} and forward search for related preprints.

Each tool design was assessed on a set of binary criteria via yes/no questions, defined to operationalise the discussed UI design trends and facets underlying \collage{} (\cref{sec:collage}). For example, the juxtaposition of voices in text entry led to the question of \textit{``Does the design afford text entry for both diegetic text and non-diegetic text?''}

The questions were developed \revised{iteratively: Top-down (deductively), the concept of \collage{} inspired question categories and first definitions of questions. Bottom-up (inductively), UI screenshots and interaction descriptions from papers inspired critical reflections on these questions, to iteratively refine the set.} For example, this clarified their definition for edge cases and added more nuance around fragmentation (e.g. the use of visual segmentation and connections).
\cref{sec:appendix_ui_analysis} lists all questions and their definitions.

\begin{table*}[]
\centering
\scriptsize
\renewcommand{\arraystretch}{1.5}
\setlength\tabcolsep{4.75pt}
\newcolumntype{L}{>{\raggedright\arraybackslash}X}
\newcolumntype{P}[1]{>{\raggedright\arraybackslash}p{#1}}
\begin{tabularx}{\textwidth}{@{}P{2cm}lP{2cm}lL@{}}
\toprule
\textbf{Framing concept} &
  \textbf{Dimension} &
  \textbf{Functional value of the fragments} &
  \textbf{Addressed user need/goal/question} &
  \textbf{Examples} \\ \midrule
Cognitive processes of writing~\cite{bhat2023suggestions, hayes2012modelingwriting} &
  Proposer &
  Getting/exploring ideas &
  I need ideas for my writing. &
  metaphor graph~\cite{kim2023metaphorian}, plot inspiration sidebar~\cite{huang2023inspo, singh2023elephant}, discussion elaboration pop-up~\cite{zhang2023visar}, idea panel~\cite{gero2022sparks}, inspirational images/audio~\cite{singh2023elephant}\\
 &
  Translator &
  Producing text &
  How can I put this idea into words? &
  text suggestions~\cite{bhat2023suggestions, dang2023choice, lee2022coauthor}, example sentences not to be used directly~\cite{shen2023convxai, shen2023convxai2}\\
 &
  Transcriber &
  Entering text &
  I need to enter these words into the system. &
  text suggestions for direct use, e.g. as sentence continuations~\cite{bhat2023suggestions, dang2023choice, jakesch2023opinionated, lee2022coauthor}\\
 &
  Evaluator &
  Evaluating text &
  Is this correct/adequate/understandable/enough/...? &
  summary cards~\cite{dang2022beyond}, highlighting/cards for tracking AI influence~\cite{chung2022talebrush, hoque2024hallmark, ito2023rewriting}\\ \midrule
Writing stages &
  Plan &
  Display/develop structure, overview &
  I need a plan/overview. &
  structured cells for prompting~\cite{biermann2022tooltocompanion, mirowski2023drama}, plot sketching canvas~\cite{chung2022talebrush}, graph of branching narrative~\cite{park2023storytelling}\\
 &
  Draft &
  Producing text &
  I need a draft. &
  text suggestions (see examples above), generation from outline graph~\cite{park2023storytelling, zhang2023visar}\\
 &
  Revision &
  Modifying text &
  I need to revise this draft. &
  movable summary cards~\cite{dang2022beyond}, sidebar for prompting AI revisions~\cite{goodman2022lampost, yuan2022wordcraft}, cells for local revisions~\cite{lin2024rambler}\\ \midrule
Beyond writing &
  Learning &
  Human learning beyond the interaction &
  I want to get better at X / learn about X. &
  highlighting/suggestions for feedback in language learning~\cite{afrin2021student, hui2023lettersmith, tsai2020lingglewrite}, filling fragments/gaps as a learning task~\cite{peng2023storyfier}\\
 &
  Organisation &
  Workflow, integration &
  How can I organise/keep track of this writing task? &
  suggestion pop-ups beyond text editor~\cite{arakawa2023catalyst}, history/timeline~\cite{afrin2021student, hoque2024hallmark} \\ \bottomrule
\end{tabularx}
\caption{\revised{Functional value of fragments in the paper set from \cref{tab:paper_set}: This table lists the purposes or roles of text fragments, when interpreting the designs in the related work through the lens of different framing concepts (first and second column). For illustration, the table also lists addressed user needs, goals, or questions, plus some examples (not exhaustive).}}
\Description{Table listing the identified functional value of fragments in the paper set with columns showing framing concepts, dimension within that concept, the corresponding functional value, and -- for illustration -- the addressed used needs, plus examples from the papers, in text. The rows are grouped into three groups by three different framing concepts: 1) Cognitive processes of writing, 2) writing stages, and 3) beyond writing.}
\label{tab:functional-value}
\end{table*}

\subsection{Collage in the UI Design of AI Writing Tools}\label{sec:analysis_results}
\cref{tab:paper_set} shows the results. The following report is structured by the shown ``\collage{} \textit{Factor}'' (i.e. number of ``yes'' answers for a design).

\subsubsection{Low -- Design choices associated with a low degree of Collage}
Designs with a lower factor stay close to traditional page UIs. They show little fragmentation and no juxtaposition of voices (i.e. no prompt writing). The user's role as the author seems largely uncontested. These designs use a single writing area, plus ``intelligent'' support, such as corrective or reflective feedback~\cite{afrin2021student, tsai2020lingglewrite}, examples~\cite{hui2023lettersmith}, sentence continuations (e.g. one~\cite{buschek2021phrasesuggestions, bhat2023suggestions, jakesch2023opinionated}, multiple~\cite{buschek2021phrasesuggestions, lee2022coauthor, lehmann2022sugglistcont}), or inspiration in a separate view~\cite{gero2022sparks, singh2023elephant}. The minimum factor is assigned to \textit{CatAlyst}~\cite{arakawa2023catalyst}, which shows a notification with a sentence suggestion to help writers return to the page from procrastination. In that UI, the only (small) aspect of \collage{} is the visual overlap of texts created by the suggestion notification. A similarly low factor appears for the system by \citet{afrin2021student}, which highlights text changes to help students reflect on how versions of their draft have evolved. Here, an aspect of \collage{} is the visual fragmentation caused by the coloured highlight boxes, which is created from multiple source texts, albeit by the same author.

\subsubsection{Medium -- Design choices associated with a medium degree of Collage}
These designs feature elements and layouts that deviate further from a predominant page, introduce more fragmentation, as well as prompting, and thus also more outside material. %
Text is shown in blocks, as cards or cells, both for AI output~\cite{dang2022beyond, peng2023storyfier} and input~\cite{lin2024rambler, mirowski2023drama}. Related, pop-ups cause visual overlap of text pieces (e.g. suggestions overlap draft~\cite{buschek2021phrasesuggestions, lee2022coauthor, dang2023choice, ito2023rewriting}), also in juxtaposed voices (e.g. prompt input overlaps draft~\cite{dang2023choice}). Further fragmentation of material from multiple sources occurs in sidebars that mix input and output~\cite{benharrak2023writerdefined, biermann2022tooltocompanion, yuan2022wordcraft}. Piece-wise text rendering or embellishments indicate sources (e.g. AI vs human~\cite{chung2022talebrush, peng2023storyfier}), AI text changes~\cite{ito2023rewriting}, and keywords~\cite{peng2023storyfier}. Finally, new interactions beyond text entry shift the user's role towards composition: writing high-level intentions (e.g. log line~\cite{mirowski2023drama}); drawing lines for story arcs~\cite{chung2022talebrush}; engaging in a conversation with the AI about the text~\cite{yuan2022wordcraft} or within the narrative world~\cite{biermann2022tooltocompanion}; or (re-)ordering text pieces, which may imply rewriting tasks that are delegated to the AI~\cite{dang2022beyond, lin2024rambler}.

\subsubsection{High -- Design choices associated with a high degree of Collage}

These designs involve rich interactions beyond typing~\cite{hoque2024hallmark, huang2023inspo, zhang2023visar}, embrace prompting~\cite{hoque2024hallmark, kim2023cellsgenlens} and material not written by the user~\cite{hoque2024hallmark, huang2023inspo, park2023storytelling, shi2023effidit}, and/or support composing on a 2D surface~\cite{kim2023cellsgenlens, lu2019inkplanner, park2023storytelling, zhang2023visar}.
In \textit{VISAR} by \citet{zhang2023visar}, AI synchronizes drafting on a page with outlining in a node-link diagram. This UI aligns exceptionally well with all four facets of \collage{} -- users arrange text \textit{fragments} in 2D space to \textit{compose} arguments, which prompts AI draft generation. Users \textit{iteratively switch voices} between this prompting and drafting, and can request \textit{outside material} as inspiration, generated by an LLM.
As another example, \textit{HaLLMark} by \citet{hoque2024hallmark} makes AI contributions transparent, with text colour plus a breakdown (\% human vs AI). It visualises prompting on a timeline and keeps past AI output as cards. As a result, users are encouraged to reflect on their acts of collage, that is, their mixing of own writing and generated fragments.
Finally, the UIs by \citet{kim2023cellsgenlens} show high degrees of \collage. For example, their copywriting tool features an extensible grid of prompt boxes and a 2D scatterplot for generated snippets. They structure such object-oriented interaction with LLMs into three components -- cells, generators, and lenses. This connects well to \collage: Cells support fragmented input; generators provide material not written by the user; and lenses on AI text shift user roles towards higher-level decision-making. This also implies writing in both voices (content and commands/prompts), as is evident from the example systems in their paper.

\subsubsection{\revised{Functional Value of Fragmentation}}
\revised{The previous sections ``counted'' aspects of \collage{}. Not every fragment has to serve the same purpose, though. Thus, here we reflect on the identified purposes or roles of text fragments. \cref{tab:functional-value} lists such functional values of the fragments, identified in the paper set (\cref{tab:paper_set}). This is structured using the cognitive processes of writing by \citet{hayes2012modelingwriting}, as adapted to writing with text suggestions by \citet{bhat2023suggestions} (i.e. proposer, translator, transcriber, evaluator). Another framing is provided by general writing stages (i.e. plan, draft, revision), and by looking for aspects beyond the writing task itself (i.e. learning, organisation). This essay provides these perspectives and interpretations as starting points for further investigation of the functional value of text fragments in such UIs, not as a definitive list.}

\begin{figure*}
    \centering
    \includegraphics[width=1\linewidth]{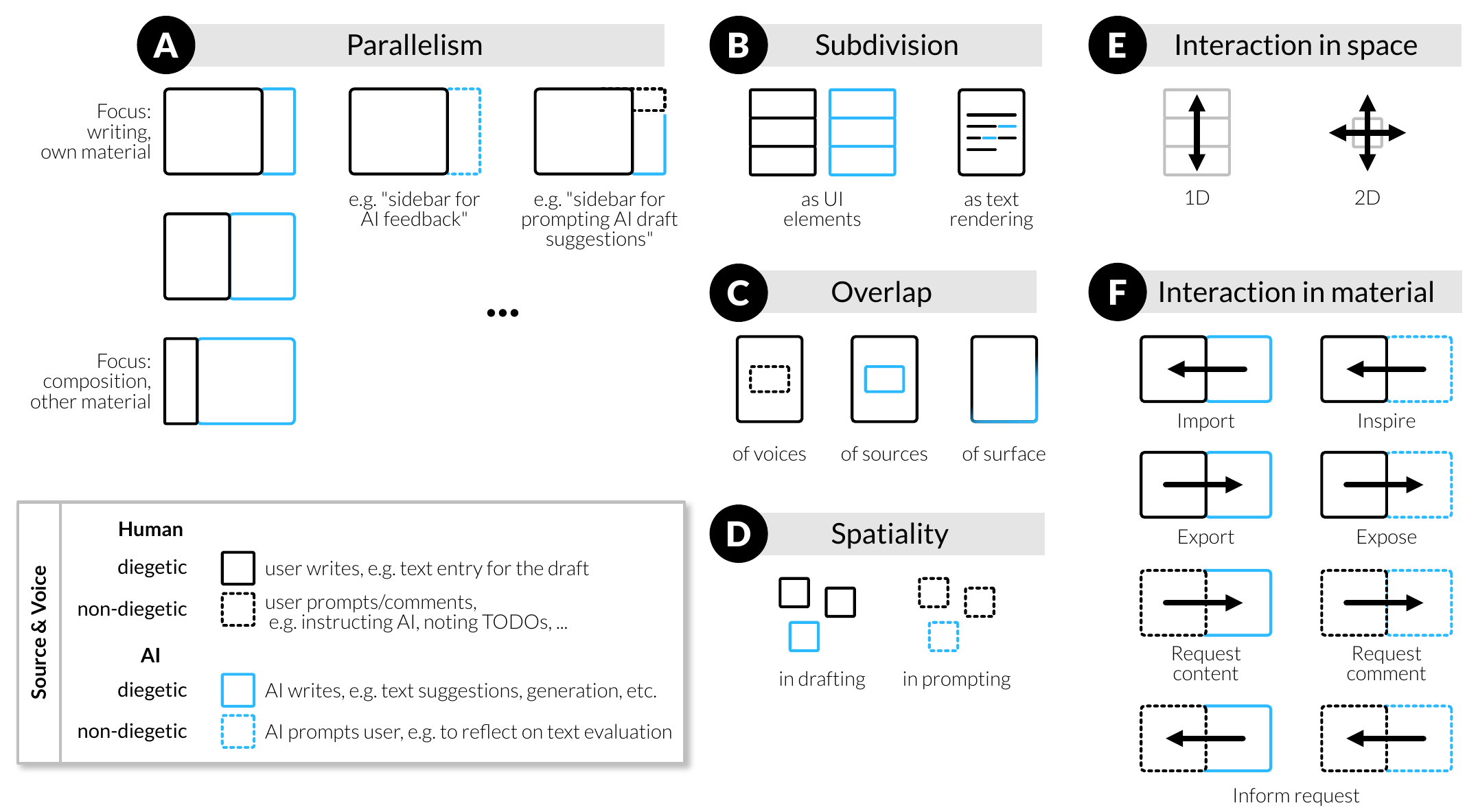}
    \caption{Using \collage{} as a lens on UI design highlights the use of different kinds of text fragments. Concretely, this figure shows the identified patterns of text display and interaction in UIs for writing with AI. We distinguish four text types (legend, bottom left), which result from combining two text sources (human vs AI, see \cref{sec:collage_sources}) and two voices (diegetic i.e. ``content text'' vs non-diegetic i.e. ``text about text'', see \cref{sec:collage_voices}). With these elements, the figure reveals four aspects of displaying text \textbf{(A, B, C, D)}, as well as two aspects of text interaction \textbf{(E, F)}. See \cref{sec:ui_patterns} for a detailed description. Best viewed in colour.}
    \Description{The figure shows the patterns described in \cref{sec:ui_patterns} as a set of small illustrations that use boxes with differently coloured lines (black: human text, blue: AI text) and line style (solid: diegetic text, dashed: non-diegetic text). Concretely, these show parallelism (different relative sizes of 2-3 such boxes), subdivision (turning one box into a grid/list of boxes, as well as rendering text pieces within one box differently), overlap (text boxes on top of each other, e.g. as a pop-up box design), spatiality (arrangement of boxes on a 2D surface beyond a usual UI layout, e.g. like a 2D canvas), interaction in space (moving text boxes), and interaction in material (8 depicted patterns, each showing two boxes, one for user text, the other for AI text, with an arrow between them; this indicates how one text relates to another, e.g. the user's draft text is processed by the AI to generate a text suggestion).}
    \label{fig:ui_patterns}
\end{figure*}

\subsection{Patterns of Text Display and Interaction in AI Writing Tools}\label{sec:ui_patterns}

The previous section discussed concrete systems as design examples. Complementary, we now extract design patterns \textit{across} individual tools. Through the lens of \collage, this means extracting text-related patterns in the UI. Specifically, this part of the essay analyses how different kinds of texts are arranged and interacted with in the UI. %
\cref{fig:ui_patterns} shows these patterns,\footnote{An example of representing a concrete UI design in this way is shown in the illustration in \cref{fig:teaser} (bottom left). It shows the pattern of text display used in the UI of \textit{HaLLMark} by \citet{hoque2024hallmark} (see Figure 1 in their paper).} described in more detail next.

\subsubsection{Four Fundamental Text Types}
As shown in \cref{fig:ui_patterns} (bottom left), we distinguish between four text types, resulting from combining two voices (\cref{sec:collage_voices}) and two sources (\cref{sec:collage_sources}):
\begin{description}
    \itemsep.5em
    \item[Text sources (colour):] Is a displayed piece of text written by the user or provided by the system/AI (e.g. generated or retrieved from an external source)?
    \item[Voices (line style):] Is a displayed piece of text diegetic (i.e. part of the content/draft) or non-diegetic (i.e. a prompt, annotation, todo, or otherwise ``writing about writing'')?
\end{description}

\subsubsection{Patterns of Text Display}
With these four text types in mind, we can read the patterns of text display, as presented in \cref{fig:ui_patterns}.

\begin{description}%
    \itemsep.5em
    \item[(A) Parallelism:]
    This patterns captures the parallel display of multiple texts. The figure indicates some examples for illustration; there are many further possibilities.
    
    \item[(B) Subdivision:]
    This pattern captures the ``splitting'' of text display into smaller parts, either as multiple UI elements (e.g. one text box per paragraph instead of a single page~\cite{lin2024rambler}) or by splitting the text rendering within one UI element (e.g. differently coloured sentences within one box~\cite{hoque2024hallmark}).
    
    \item[(C) Overlap:]
    In this pattern, the same screen area is used for several texts such that they visually overlap. This includes both overlapping voices (e.g. a pop-up prompt box overlaps the writing area~\cite{dang2023choice}) and sources (e.g. a pop-up with text suggestions overlaps the writing area~\cite{lee2022coauthor}). It also includes a third kind of overlap, in which a single visual writing surface serves multiple text types (e.g. a page on which both user and AI can co-write without further distinction).
    
    \item[(D) Spatiality:]
    This captures the use of spatial arrangements of text displays in the UI, including in 2D, such as canvas-like UIs, mind maps, or other graph-like representations.
\end{description}

\begin{figure*}
    \centering
    \includegraphics[width=\linewidth]{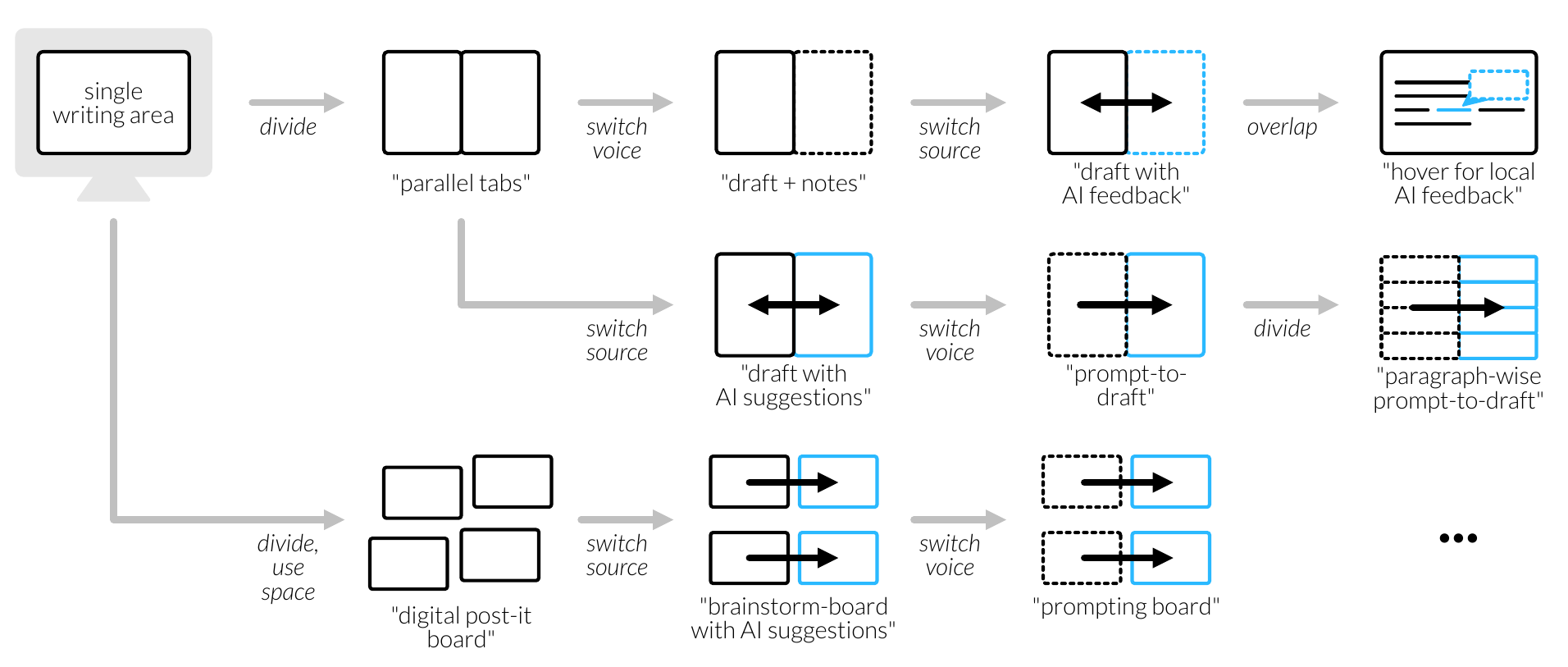}
    \caption{Illustration of using the patterns (\cref{fig:ui_patterns}) to construct new tool designs. Starting from a traditional single writing area (top left), we morph the text fragments displayed in the UI into other (sub)divisions, arrangements, voices, and sources. The arrows and action words in between the designs indicate what is changed, while the ``titles'' below each design shortly describe a tool idea consistent with this way of displaying text. Black arrows on the designs indicate the (predominant) intended relationship between the text fragments (cf. \cref{fig:ui_patterns}D); other relationships might also be possible. See \cref{sec:ui_morphing} for details and \cref{sec:appendix_example_morphs} for descriptions of all shown design ideas.}
    \Description{Figure shows UI design examples represented in the boxes-style introduced earlier in \cref{fig:ui_patterns}. The concrete examples are linked by arrows, which are labelled by ``collage actions'', such as ``divide'' when a new design is derived from the previous one by splitting a text view into multiple ones. In this way, the figure ``evolves'' a single page UI into ten new ideas of increasing complexity. Detailed textual descriptions of these ten ideas are available in \cref{sec:appendix_example_morphs}.}
    \label{fig:ui_morphing}
\end{figure*}

\subsubsection{Patterns of Text Interaction}
The remaining parts of \cref{fig:ui_patterns} show the patterns of how text is referred to throughout interaction. 

\begin{description}%
    \itemsep.5em
    \item[(E) Interaction in space:]
    This describes the pattern of enabling users to change the spatial configuration of text displayed in the UI. For example, in a 1D case, this might be reordering paragraphs via drag and drop~\cite{dang2022beyond, lin2024rambler}, while in 2D, users might move pieces on a canvas~\cite{zhang2023visar}.
    
    \item[(F) Interaction in material:]
    Throughout interaction, the four text types are leveraged in various typical ways in relation to each other. %
    These are not mutually exclusive and may often be chained or go hand-in-hand (e.g. import + export). %

    \vspace{.5em}
    \begin{itemize}
        \itemsep.5em
        \item \textit{Import:} The user's text is influenced by AI-provided diegetic text. For example, the user accepts a sentence continuation for the draft or is influenced by reading it~\cite{buschek2021phrasesuggestions, jakesch2023opinionated, lee2022coauthor}.
        \item \textit{Export:} This is the counterpart to \textit{Import}. AI processes the user's writing to provide diegetic text (e.g. the typical case of suggestions on how to continue the draft).
        \item \textit{Inspire:} The user's text is influenced by AI-provided non-diegetic text. For example, the AI generates feedback comments on the user's draft to prompt human reflection and subsequent revisions~\cite{benharrak2023writerdefined}.
        \item \textit{Expose:} This is the counterpart to \textit{Inspire}. The AI processes the user's writing in order to provide non-diegetic text, for example, to generate feedback on the text.
        \item \textit{Request content:} The user's non-diegetic text is processed by the AI in order to create diegetic text. For example, the user enters a prompt to instruct the model to generate draft text (e.g. ``Write a blog post on topic X'').
        \item \textit{Request comment:} The user's non-diegetic text is processed by the AI to create non-diegetic text. For example, the user asks the AI for feedback or inspiration, not for a draft text directly (e.g. ``How could I make my introduction more engaging for young readers?'').
        \item \textit{Inform request:} This is the counterpart to the \textit{requests} above. The user picks up on AI-generated text to write or revise the (next) comment or prompt text.
    \end{itemize}

\end{description}

\section{Constructive Lens: Leveraging \collage{} in UI Design}\label{sec:constructive_lens}

This part employs \collage{} as a constructive lens to inspire new UI designs for AI writing tools. We start with a bottom-up approach using the patterns (\cref{sec:ui_patterns}), followed by two broader design explorations.

\subsection{Evolving the Page UI via \collage{} Patterns}\label{sec:ui_morphing}

\begin{figure*}
    \centering
    \includegraphics[width=0.75\linewidth]{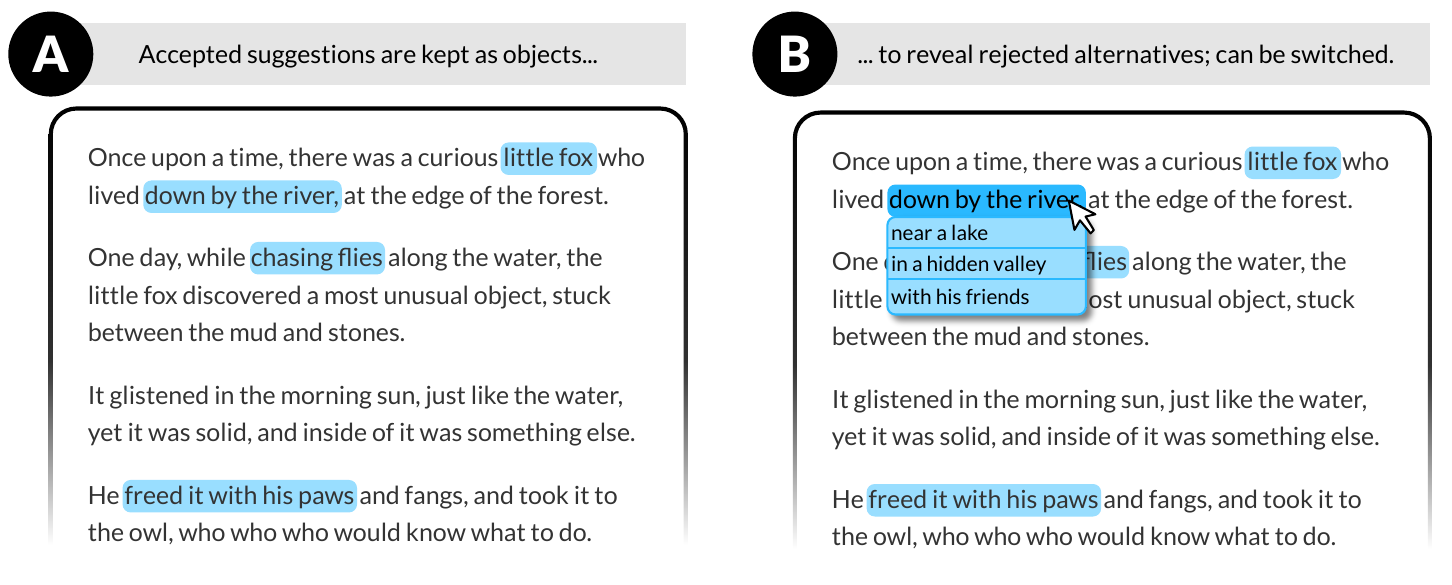}
    \caption{Inspired by text fragmentation in \collage{}, this UI design idea \textbf{(A)} keeps and shows the seams between text fragments written by the user (black) and those suggested by the AI (blue). In this way, \textbf{the UI turns the user's past suggestion choices into interactive objects: Clicking on one such object \textbf{(B)} reveals the alternative suggestions that the user rejected back then.} This design facilitates transparency around human vs AI writing \revised{(e.g. to allow writers to double-check these AI-provided text pieces for accidental plagiarism)}, may help communicate one writer's intention to co-authors in collaborative writing, and could address ``agony of choice'' situations if the user likes multiple suggested options, as no option is lost. Users can also retroactively change their mind by selecting a different alternative. In a more advanced tool design, this could prompt an AI-rewrite of the subsequent story parts, to match the semantics of the updated suggestion choice. See \cref{sec:design_exploration1_suggestions} for a detailed description.}
    \Description{The figure shows the UI idea described in the caption with two UI mockups: The left mockup shows an example text on a page view with some words or sentence-parts highlighted in blue. The right mockup shows the same UI in a different state: Here, the user is hovering over one highlighted text part with the mouse cursor. This has opened a pop-up list directly below the text snippet, which displays the alternative text suggestions shown by the AI at the time in the past, when the user had selected the one they are hovering over now.}
    \label{fig:design_exploration1}
\end{figure*}

As a first step towards constructive use of \collage{}, we consider the patterns from \cref{sec:analysis_results} (\cref{fig:ui_patterns}). They all relate to \textit{text elements} in the UI. This is different to more general UI patterns that might involve layouting various kinds of input and output elements, such as sliders, buttons, panels, images, labels, and so on. 
A consequence of this focus is that this representation of a UI facilitates counterfactual thinking~\cite{oulasvirta2022counterfactual} specifically around the use of text. For a UI design expressed in this way, we can ask: 
\begin{itemize}
    \item \textbf{Fragmentation:} \textit{What would the interaction be like if element X was divided, overlapped with Y, etc.?}
    \item \textbf{Sources:} \textit{What would the interaction be like if element X was text of a different source?}
    \item \textbf{Voices:} \textit{What  would the interaction be like if element X was text in a different voice?}
\end{itemize}

In this way, we employ these patterns for a kind of morphological analysis~\cite{card1991morphological} to generate new designs. These are not full UI designs but they indicate which types of texts are shown, in which way these are fragmented, and how the texts in these fragments (abstractly) influence each other. 
\cref{fig:ui_morphing} illustrates this process with a set of example design ideas, created in this way. In addition, \cref{sec:appendix_example_morphs} provides detailed descriptions of these examples.

This rather ``mechanical'' application of the patterns of \collage{} is a bottom-up approach of coming up with new tool designs. Complementary, in the following two design explorations, we demonstrate a top-down approach. That is, we derive design inspiration from considering \collage{} in the context of other HCI concepts and theory.

\subsection{Design Exploration 1: Embracing \collage{} for Reification of Text-related Decision-making}\label{sec:design_exploration1}

Imagine cutting text snippets from a magazine to compose a poem. In this design exploration, we embrace that literary collage is comfortable with 1) showing the ``seams'' of text snippets, and 2) with keeping and treating text pieces as objects.

What does this mean for a digital writing tool with AI?
If we keep a visible, interactive seam between text pieces, this amounts to a UI representation of past collage actions, or, in other words, their \textit{reification}~\cite{beaudouin2000reification}. 
Concretely, reification turns abstract concepts into interactive objects in the UI.
This design direction further relates to \textit{seamfulness}, which research has explored also in relation to visible histories of past actions (see the review by \citet{inman2019seams}).

\subsubsection{Reification of the User's Decision Moments about Text Suggestions}\label{sec:design_exploration1_suggestions}

\begin{figure*}
    \centering
    \includegraphics[width=0.75\linewidth]{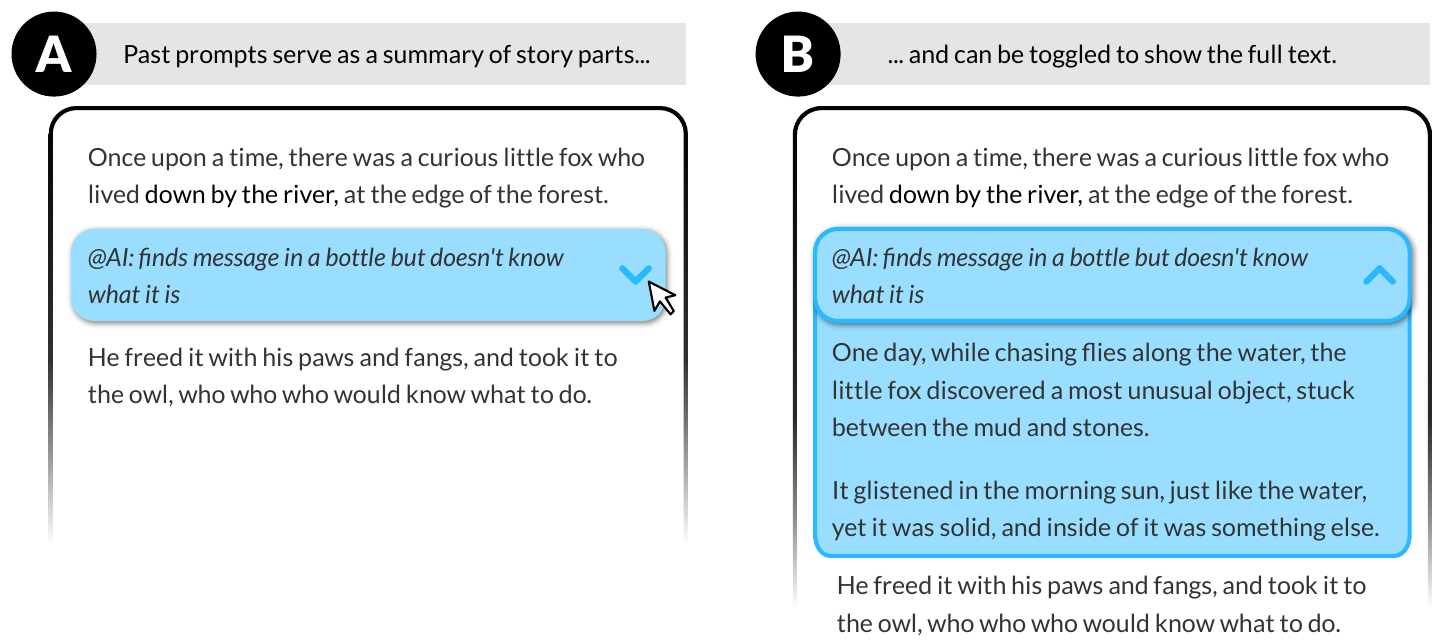}
    \caption{Inspired by the juxtaposition of text from different voices in \collage{}, this design idea \textbf{(A)} shows the seams between text fragments written as part of the draft (black) and those entered to prompt the AI (blue). In this way, \textbf{the UI turns the user's past prompts into interactive objects: Clicking on one such object \textbf{(B)} reveals the corresponding text generated by the AI} (which the user may also edit just as any other text). This design allows users to further benefit from the effort they put into their prompts, since prompt texts do not disappear. Instead, the prompt texts serve as summaries of the respective parts of the draft. This might be useful to gain an overview when coming back to a writing project or when going through the latest version shared by a collaborator. It might also facilitate efficient context understanding for editing tasks, without having to read everything. See \cref{sec:design_exploration1_prompts} for a detailed description.}
    \Description{The figure shows the UI idea described in the caption with two UI mockups: The left mockup shows an example text on a page view with three paragraphs. The first and third one are normal text of the story. The second paragraph is a blue card-like view that contains a prompt written to the UI. The right mockup shows the same UI in a different state: Here, the user has clicked on the prompt card, which has expanded it. Now, it shows the generated story text beneath the prompt.}
    \label{fig:design_exploration1a}
\end{figure*}

We first explore reification of suggestion choices. Current AI writing tools that suggest text (e.g.~\cite{buschek2021phrasesuggestions, lee2022coauthor, jakesch2023opinionated}) assume that the user will ``accept'' or ``reject'' a suggestion (e.g. by pressing a key, choosing from a list, or by ignoring it altogether). 
Crucially, after acceptance, the draft no longer reveals this moment of choice. While some recent designs use colour to track human vs AI text (e.g.~\cite{hoque2024hallmark}), this highlighting is mostly a ``passive'' object, in that it informs the viewer but does not afford actions that affect the text.

This need not be the case. Inspired by \collage, we envision an alternative design direction that reifies the user's editorial decisions around accepting/rejecting a text fragment (i.e. an act of collage).

\cref{fig:design_exploration1} shows what such a UI might look like: Here, the system provides sentence continuations, which the user can accept or ignore. If accepted, the inserted text fragment is formatted to ``reveal its seams'' using a coloured box. If the user clicks on this text fragment, they can see the list of rejected alternative suggestions, originally displayed at the moment of choice. They can also actively change their choice by clicking on a new suggestion in that list. In a more advanced version, this change might prompt an AI-rewrite of the subsequent parts, to semantically fit to the updated choice.

Such a design might offer compelling benefits: \revised{Highlighting accepted suggestions} provides some transparency about sources, \revised{for example to allow writers to check that AI-provided text does not introduce plagiarism. It also provides transparency} about decision-making. Seeing what was not chosen could help communicate one writer's intention to others in collaborative writing software. It could also counter potential ``agony of choice'' situations when writers like multiple suggested options -- no suggested text is lost, they can always come back and change their mind. On the flip side, having this possibility to return to earlier decisions might distract writers from moving on with the draft. Future work could investigate the impact of such a design empirically.  %

\subsubsection{Extension: Fragmented Document History}
\revised{The idea outlined above (and in \cref{fig:design_exploration1}) does not reveal what happens when users manually edit suggested text. In the spirit of this design exploration, we could introduce more fragmentation: Upon editing suggested text, the unedited version could automatically be added as an option to the dropdown list.}

\revised{More generally, we could introduce a ``tracked fragment mode'' which allows users to turn \textit{any} selected text into an object that gathers all its further evolving versions in a dropdown list or some other adequate selection view. This is different from existing ``track changes'' features, which typically only support accept/reject, and then \textit{remove} the associated interactive object. It is also different from a document version history, which is typically \textit{global}. In contrast, this idea here provides a fragmented document history: It supports \textit{repeated local switching} of text fragments, which might be useful for exploration.}

\subsubsection{Reification of the User's Prompt Texts}\label{sec:design_exploration1_prompts}

A variation on the design direction is to reify the user's \textit{prompting moments}. For this, we assume an AI writing tool that the user can prompt to draft a paragraph (e.g. ``Continue the story such that Alice meets Bob at the supermarket''). 
Current designs of this kind assume that the prompt is ephemeral \revised{and disappears after ``submitting'' it. Even in a chat-like UI, such as} in ChatGPT, previous prompts will eventually scroll out of view as the chat history grows. 

Instead, here we envision a UI that keeps prompts as persistent fragments to be leveraged as a summary. \cref{fig:design_exploration1a} illustrates this.

The figure shows an AI tool that allows the user to enter prompts directly on a page view. These might be marked by a keyword, such as ``@AI: ...'' (e.g. as in the mockups by~\citet{yang2019sketchingnlp}).
The AI responds by directly adding its generated text below this prompt. 
The prompt and this response are reified as one object.
Similar to the first design direction, this could enable editing the prompt later on, to update the generated text.

Beyond that feature, this object can be collapsed to only display the user's prompt text -- or it can be expanded to show the generated text (which the user might also have edited manually later on).
In this way, the prompt can be reused as a kind of summary. This might be useful to gain an overview and context understanding for local edits in longer drafts, without having to read through a ``wall of text''. Supporting this idea, collapsing text is well known from writing code. For instance, many IDEs allow users to collapse functions/methods such that only the signature is shown but not their body. Moreover, this feature might also support coming back to a writing project after a longer break, or might help with catching up with the latest version shared by a collaborator.

\begin{figure*}
    \centering
    \includegraphics[width=0.75\linewidth]{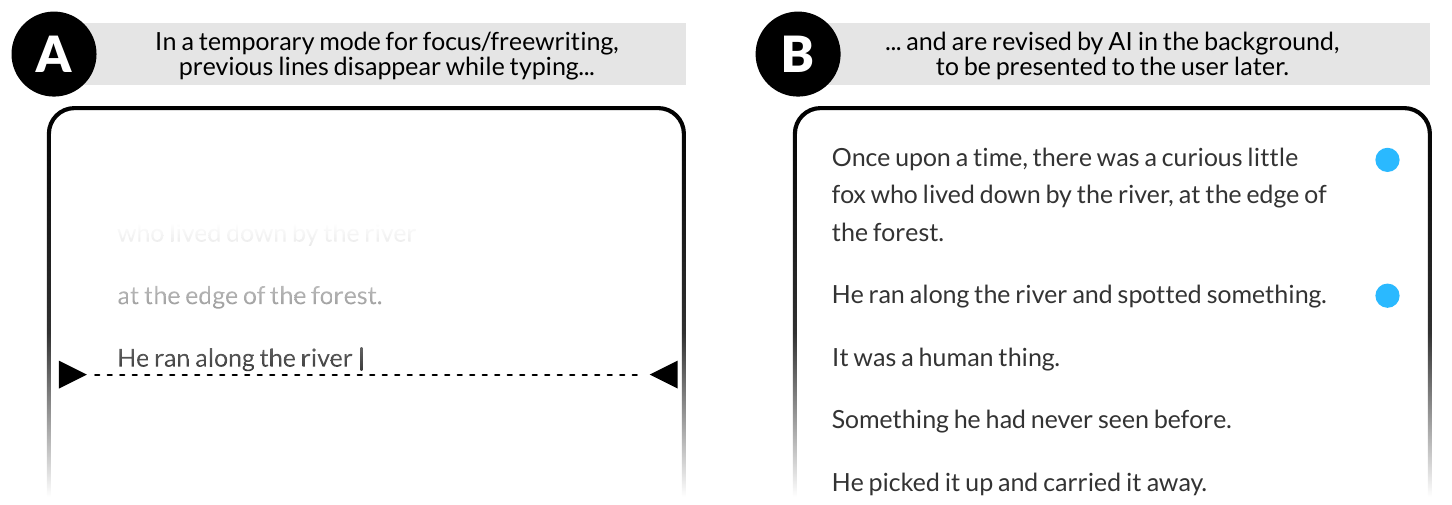}
    \caption{Inspired by \textit{rejecting} \collage{}, \textbf{this design idea \textbf{(A)} aims to minimise the number of text fragments displayed at a time}. The result is a ``focus mode''', in which previously entered lines disappear while typing (e.g. akin to submitting a line in a chat app UI, which scrolls earlier messages out of view). In line with the ``freewriting'' method~\cite{li2007freewriting}, this might help users with pushing on with the draft, instead of perfecting their wording or otherwise getting tempted by (premature) revisions. \textbf{AI processes the user's sentences in the background} to fix typos and grammar, remove redundancies, extend fragments into complete sentences, and so on. In this way, the user receives a cleaned-up version when leaving this writing mode \textbf{(B)}. Lines with AI changes could be marked (e.g. by blue dots) to provide transparency about AI revisions, without visually fragmenting the text display with word-based highlights, as was the case in \cref{fig:design_exploration1}. See \cref{sec:design_exploration2} for a detailed description.}
    \Description{The figure shows the UI idea described in the caption with two UI mockups: The left mockup shows an example text on a page view. Only one line is fully visible, the previous lines are faded by rendering them in a less dark shade of grey. In this way, no more than the last line is really still readable. The current line of text with the user's caret is underlined. The right mockup shows a page view after the user has finished the focus session: Here the full text is revealed again, with some AI changes. Lines with AI changes are marked by a small blue dot in the right margin of the page.}
    \label{fig:design_exploration2}
\end{figure*}

\subsection{Design Exploration 2: Rejecting \collage{} for Focused Writing UIs}\label{sec:design_exploration2}

Now we turn the perspective on its head by \textit{rejecting} \collage. In other words, what if the design goal is to keep fragmentation low, while still allowing users to benefit from AI support?

In terms of the identified patterns (\cref{sec:ui_patterns}), this means minimising aspects such as parallelism, subdivision, overlap, and spatiality.
For this design exploration, we thus ask: What if there was a writing UI that only shows a \textit{single piece of text} at a time? 

Concretely, here we consider a single sentence. \cref{fig:design_exploration2} (A) shows such a design. As the user enters a sentence, previous lines disappear (e.g. as in a chat conversation UI, where previous messages eventually scroll out of view).
At first glance, the downsides are plentiful: No context, no way of revising earlier parts, very hard to resume writing, and so on.

However, this design direction might be appealing as a dedicated ``focus mode'': Writing line by line frees the writer from getting distracted by earlier imperfections and the temptation to revise that digital word processors have introduced (cf.~\cite{elbow1998freewriting, joram1992revise}). 
Similarly, it is intended to nudge the user to not think too much about the exact wording, as that is quickly removed from the screen anyway. What remains is the writer's mental model of the narrative -- which starts with ideas, not precise wording~\cite{flower1981writingprocess}.

Together, these aspects align well with the writing technique of \textit{freewriting}, which is about ``writing without stopping and editing''~\cite{li2007freewriting}. The goal is to keep writing, even if no great ideas come to mind, potentially even just repeating the same words multiple times, until inspiration strikes again. It is not intended as a prolonged activity but as a remedy in situations of ``writer's block''.

In a design like this, AI remains largely in the background, to not introduce its ``voice'' into the UI during drafting. As presented in \cref{fig:design_exploration2} (B), the AI processes the user's sentences (including the ones that have disappeared from the screen) to revise them, such that the user ends up with a cleaned version when leaving the ``focus mode''. For example, the AI might be used to fix typos and grammar, remove redundancies, extend fragments into complete sentences, or make other edits that improve the overall coherence.

Finally, the user is presented with this revised version of their text from the focus session. They can edit it manually, use it partly or directly in their further draft, and so on. Many design variants are possible here, such as toggling between the raw outcome and the AI-revised version, switching certain types of AI edits on or off, etc. Our mockup in \cref{fig:design_exploration2} (B) illustrates a minimal design, which aims to avoid fragmentation in this step as well. For instance, we only indicate AI revisions per line (blue dots), not per word.

\section{Critical Lens: Concerns and Collage Then and Now}\label{sec:critical_lens}

If recent AI tools successfully address user needs, and we can locate aspects of collage in their UIs, we could see collage as a \textit{remedy}. However, literary collage was also used to surface problems and challenges, such as by the writers of literary Expressionism. That is, collage was used to express \textit{symptoms}. Here we explore this view. Concretely, this part of the essay reflects on issues around writing with AI by engaging with the concerns that the writers of literary Expressionism historically articulated through literary collage. 

In brief, literary Expressionism was a literary movement in Germany from around 1910 to 1925 and is considered an important part of avant-garde~\cite{bogner2005einfuehrung}: It introduced new aesthetics to react to the drastic societal and cultural changes of the time. This reaction was typically negative, reflecting in its stylistic devices what it sought to criticise. Core themes include mechanisation and industrialisation, life as an individual in growing cities, and the impact of technology with the establishment of new mass media structures.

\subsection{Treating Language as Prefabricated Material}

Expressionist writers used literary collages to dissolve structure. Not limited by meaningfully connecting sentences, they juxtaposed fragments from different sources and on different topics. 
They thus treated language as ``prefabricated material''~\cite{vietta1974}: For instance, this mirrored how they saw the relationship of the press and the urban individual, who in thinking about the world cannot escape referencing others' writing (e.g. the newspaper). %

Through collage, they further expressed overwhelming feelings of discontinuity, dissociation, simultaneity of events and the ``acceleration'' of life in growing industrialised environments~\cite{anz2016litexpr, bogner2005einfuehrung, vietta1974}.
Their collages picked up phrases and clichés from everyday language in circulation, as ``stereotypical and interchangeable quotations''~\cite{vietta1974}. In using these, they also criticised a ``template-like'' way of communication~\cite{bogner2005einfuehrung}.

This essay proposes that this critical perspective translates well to writing with AI today:

First, LLMs are also trained on language in circulation, that is, text from the internet.
Second, AI writing tools produce prefabricated language material for people to use, such as through text suggestions, including as templates for communication (e.g.~\cite{chen2019smartcompose, kannan2016smartreply, robertson2021cannotrelpy}).
Third, this indeed produces clichés and stereotypical language.
For example, \citet{arnold2020predictable} found that writing with word suggestions makes text more predictable. Similarly, writing with LLMs can make texts less diverse~\cite{padmakumar2023contentdiversity}. Other work~\cite{buschek2021phrasesuggestions} noted an increase of ``business phrases'' from the training data of the LLM when writing emails with text suggestions. %

In summary, we can now prefabricate language material for users to an extent that was unimaginable at the time of the Expressionists. Moreover, with LLMs, this linguistic material has become flexible, pliable, as fluid as fluent sentences that fit into any context we desire. LLMs are trained to extend sentences, to fill in gaps -- they bring both the material and the glue.

So far, at least, this seems to come at the risk of making it easy to construct more similar things.

\textbf{For design,} this means that we need to consider carefully how prefabricated text material is made actionable in the UI of AI tools. Even small details may matter. For example, related work reported that a button to directly integrate AI summaries made people think of these as text suggestions, although they were intended as material for reflection~\cite{dang2022beyond}. Finally, tool design needs to be either specific or adaptable, since ``template-like'' language may be useful in some use-cases while undesirable in others (e.g. contrast these designs for legal writing~\cite{han20920textlets} vs creative writing~\cite{singh2023elephant}). %

\subsection{Language Scarcity}

If AI provides language as prefabricated material, then AI in writing has so far been about writing \textit{less}, at least manually.
Fittingly, Expressionist writing also used ``language scarcity''. 

Their engagement with the ``shock of modernisation'', which speeds up life, shows in their ``will to focus on the essential''~\cite{anz2016litexpr}: The condensed style of chaining fragments (\textit{parataxis}) is intended to evoke a sense of speed in the reader. At the same time, writing shorter pieces itself fits into a fast life that leaves little time for writing longer forms.

\citet{anz2016litexpr} reports an illustration of this stylistic device: ``The trees and flowers bloom'' is shortened to ``Tree and flower blooms'', which is distilled to ``Tree blooms flower'' to finally arrive at ``Flowering''\footnote{In the German original, this final line says ``Blüte'' (flowering/blossom/bloom).}.

In the context of AI tools, this looks a lot like prompt engineering.

For instance, consider prompts for text generators or text-to-image generators, as in systems like ChatGPT. Here, many users express their desired results in a fragmented style\footnote{E.g. see typical text-to-image prompts shared online, such as on reddit or dedicated prompt exchange websites.} to distil a maximum of meaning into a minimum of text. In other words, both AI prompting and Expressionistic parataxis take shortcuts to express more with less, by recruiting associations based on prior exposure (human: life, LLM: training data).

Beyond prompting, text entry in general has a long tradition in HCI research with a focus on speed and keystroke savings~\cite{kristensson2014inviscid}, and how these measures are impacted by ``intelligent'' features, such as autocorrection and word suggestions~\cite{banovic2019expertentry, lehmann2023morethanspeed}.

In summary, distilled delegation via prompts, plus system edits and suggestions, promise us to get more done faster, with fewer actions and less effort. 
However, in contrast to writing input to generative AI, the Expressionists distilled text to process their experience of the world and prompt associations in \textit{human} readers.

\textbf{For design,} this leaves us with critical questions: Which use-cases benefit from writing less ourselves? When is writing about ``getting text''? And when is it about working through \textit{our} experiences and associations, that is, thinking?

\subsection{Experiencing Linguistic Worlds with an Agenda}

Expressionist writers also dealt with the theme of thinking being replaced by produced language. %
Mass media innovations served a new mix of knowledge and message, creating an overwhelming state of simultaneous inputs and influences~\cite{anz2016litexpr, bogner2005einfuehrung}. %
They warned that this replaces people's own ``primary experiences'' with language surrogates, such as when reading about an event in the news~\cite{vietta1974}.

Today, we might look back and say that this is true for any old new medium, including the internet. Nevertheless, this lens highlights a critical angle for writing with AI: Newspapers of the 1910s, just as blog posts 100 years later, have always been written by people. In contrast, with generative AI, we move towards a world of media that no longer requires human primary experience -- not even on the creator's side.

Related, literary collage in Expressionism mirrored what it sought to criticise -- a broken relationship of subject, language, and reality~\cite{vietta1974}: Language was no longer tied to human experience of the world but had become an object \textit{in} the world. The Expressionists referred to slogans, signals and signs encountered in growing cities. Noticeably, these are all language objects that aim to convince, sell, command, and suggest something to the urban citizen.

This relates to AI tools in two ways: 
First, language is turned into (digital) objects in the UI, including in the UI ideas in this essay (\cref{sec:constructive_lens}). If a user intends to write about something they care about in the world, they now also have to interact with this ``other world'' of text objects while doing so. Put simply: Text suggestions and other such objects may draw in or distract the writer.
Indeed, this has been observed in HCI research. For example, studies report that some participants experienced distraction by suggestions~\cite{buschek2021phrasesuggestions} or examined this for different suggestion displays~\cite{dang2023choice}. \citet{bhat2023suggestions} extended the traditional cognitive process model of writing~\cite{flower1981writingprocess} to reflect the cognitive engagement with suggestions, based on their study of how people think about text suggestions during writing.%

Second, the idea of a world of linguistic objects designed to influence us fits well with LLMs as a ``world model'' trained on text, including material that expresses opinions. Indeed, related work found that LLMs have biases~\cite{bender2021parrots}, favour some views over others~\cite{johnson2022americanghost}, and -- through interaction -- may influence both writing and writers~\cite{jakesch2023opinionated}. Finally, the critique of a broken relationship between language and reality fits to LLMs' ``hallucinations''~\cite{ji2023hallucinationsurvey}.

In conclusion, we can now create ``linguistic worlds'' that fabricate words faster, without a writer, and their outputs are at once disconnected from the real world and affecting it. The Expressionists would be shocked. %

\textbf{For design,} this means that UIs of AI writing tools could be conceptualised as a boundary between worlds -- the user's world of experience and the system's text world, from which material is fabricated. This perspective highlights, for example, that concrete design choices could facilitate exchanges between these worlds, as in human-AI co-writing, in which case distraction vs positive integration appear as key challenges. Alternatively, UI designs could favour controlled exchanges, in which case the user needs to be supported to control or audit AI-generated material.

\subsection{Summary}

With this critical lens, we have looked at three broader issues around writing with AI: 1) writing more ``standardised'' material, 2) writing less and thus losing out on writing as a thinking tool, and 3) getting distracted by or drawn into (opinionated) worlds of language objects.
While individual concerns have been brought up before, this essay added new connections and thus hopefully stimulating perspectives.

Finally, the DIS'24 conference theme asks ``Why Design?''\footnote{\url{https://dis.acm.org/2024/}} and points to both design as an optimistic endeavour, as well as to its potential limitations in uncertain times.
In this light, the critical lens in this essay seeks out ``Why''s beyond the technological capabilities (i.e. ``Why? Because we can''), by reflecting on the ``Why''s of past innovators of writing. As revealed, they used forms that can be rediscovered today, in facets of the UI design of AI writing tools. Crucially, they used these forms to write responses to the uncertainties of their time, some of which seem to align with challenges today, perhaps (un)surprisingly well.

\section{Conclusion}

New writing tools and technology impact how we work with text~\cite{kirschenbaum2016track} but that is not the only source of innovation in writing. Historically, new writing forms have emerged also from developments in culture and art, and as a critical reaction to societal changes. Here we have explored one particular literary form -- collage -- and how it might be transferred to the context of writing with AI tools. %

Zooming out, it is a pattern in HCI that emerging technological capabilities, such as recently LLMs, inspire new interactive systems and tools. For AI writing tools, the next step is to foster connections between insights from individual design instances, in order to develop a rich body of intermediate-level concepts~\cite{hook2012strongconcepts} and theory~\cite{beaudouin2021gentheory}. As pointed out by \citet{hook2012strongconcepts}, this is a discursive process. As such, it likely benefits from diverse starting points and perspectives and is not achieved in a single paper but throughout a conversation of many researchers and their contributions. 

An essay, as a literary form, is an ``attempt''\footnote{\url{https://www.etymonline.com/word/essay}} -- in this case, to start from historical writing innovation to develop a perspective on recent innovations of writing tools.
The resulting concept of \collage{} in the design of AI writing tools highlights the fragmentation of UIs and the writing process therein. This provides a cross-cutting concept that can be practically located in UI design to stimulate future discourse along analytical, constructive, and critical dimensions.

\begin{acks}
The author thanks Florian Lehmann and Xiaodan Tian for discussions around freewriting and AI, and the anonymous DIS'24 reviewers for their useful feedback and suggestions, including the one on sharing the fragmentation of the writing process for this essay (\cref{sec:appendix_meta}). Daniel Buschek is supported by a Google Research Scholar Award. This project is partly funded by the Bavarian State Ministry of Science and the Arts and coordinated by the Bavarian Research Institute for Digital Transformation (bidt).
\end{acks}

\bibliographystyle{ACM-Reference-Format}
\bibliography{bibliography}

\appendix
\section{UI Analysis}\label{sec:appendix_ui_analysis}
Here we describe the questions used in the paper analysis in \cref{sec:analytical_lens}. The question codes (bold) correspond with those in the question columns in \cref{tab:paper_set}.

\subsection{Fragmentation}
Questions related to the facet of \textit{fragmentation} (\cref{sec:collage_fragmentation}):
\begin{itemize}
    \itemsep.5em
    \item \textbf{In:} \textit{Does the UI show multiple UI elements for text entry in parallel?} 
    \item \textbf{Out:} \textit{Does the UI show multiple AI-generated/processed text pieces in parallel?} 
    \item \textbf{Spatial:} \textit{Does the UI afford or use the spatial arrangement of text input and/or output?} Since any text needs to be \textit{somewhere} on the screen, this is only a ``yes'' if the arrangement goes beyond just layouting the UI. Typical positive examples of this are canvas views or graph-like presentations, but also lists of text cards.
    \item \textbf{Segmentation:} \textit{Does the UI visually delineate pieces of text material as visual objects?} Positive examples include text snippets presented in cards or highlighting parts of text. In contrast, a simple line break or the whitespace between lines is not enough for a ``yes'' here.
    \item \textbf{Connection:} \textit{Does the UI visually connect pieces of text material?} Positive examples include connecting lines (e.g. in a graph-like view) but also parallel highlighting of multiple text pieces (e.g. to show that they are somehow related).
\end{itemize}

\subsection{Voices}
Questions related to the \textit{juxtaposition of voices} (\cref{sec:collage_voices}):
\begin{itemize}
    \item \textbf{Diegetic + non-diegetic:} \textit{Does the UI afford text entry for both voices?} Yes, if the UI allows or even demands that the user enters text for both the content/draft (i.e. diegetic writing), as well as for prompting the AI, taking notes, or otherwise text beyond the content/draft (i.e. non-diegetic writing).
\end{itemize}

\subsection{Sources}
Questions related to the integration of \textit{text from multiple sources} (\cref{sec:collage_sources}):
\begin{itemize}
    \itemsep.5em
    \item \textbf{Provides text:} \textit{Does the UI show generated or retrieved text material?} Yes, if the UI shows generated or retrieved text material that can be expected to bring in phrasing, wording and/or information not entered by the user (e.g. not just translation, not just error correction).
    \item \textbf{Indicator:} \textit{Does the UI indicate the source of text material?} Yes, if the UI has \textit{dedicated} elements for sources. This might be on the level of ``human vs AI'' or on the level of specific references. Note that this is not automatically a ``yes'' just because users could infer the source (e.g. from a layout with human text on the left, AI text on the right) -- there needs to be a visual element or explaining text specifically added as a source indicator.
\end{itemize}

\subsection{User Role}
Questions related to \textit{shifting the users' roles} (\cref{sec:collage_roles}):
\begin{itemize}
    \itemsep.5em
    \item \textbf{Explore:} \textit{Does the UI offer elements for exploration of text material not written by the user?} Yes, if the UI has \textit{dedicated} elements for exploration of text material that is not written by the user (e.g. search-like features, filters, text ranking/scores, history of generations); not just a ``refresh'' button.
    \item \textbf{Arrange:} \textit{Does the UI afford interactions with the spatial arrangement of text pieces?} Yes, if the UI afford interactions to directly manipulate or indirectly influence the linear order or spatial arrangement of text pieces. This needs to be something more than normal text area functionality (e.g. not just rearranging paragraphs in a text box by copy/pasting them).
    \item \textbf{Curate:} \textit{Does the UI afford user choice w.r.t. AI-provided text?} Yes, if the UI affords user choice of either 1) accepting vs rejecting text, or 2) selecting text among a set of options. This needs to be supported with \textit{dedicated} interactions or UI elements; not just a general capability to read and/or copy/paste or edit AI-provided text.
\end{itemize}

\section{Evolving the Page UI: Examples}\label{sec:appendix_example_morphs}

Here we describe the design ideas given as examples in \cref{fig:ui_morphing} in more detail. We list these examples in their order of appearance in that figure, when read from left-to-right, top-to-bottom. Note that these are concrete examples of writing tool concepts in line with the respective text displays depicted in \cref{fig:ui_morphing} -- but they are \textit{not the only} tool ideas to fit these depictions. That is, more than one tool concept can be realised with the same number, arrangement, etc. of text elements.

\subsection{``Parallel tabs''}
In this idea, the UI allows users to open two or more parallel views on the same text file. In this way, users can work on two parts of a longer document without having to scroll back-and-forth repeatedly. Alternatively, they might only work on them yet require to view the other one for context. For example, this might be useful when writing a scientific paper, to work on both the intro and conclusion in order to achieve a coherent framing that ``closes the circle''. This design does not involve AI and is intended as a simple first example idea here. There are products that already offer this feature, such as \textit{Obsidian}\footnote{\url{https://obsidian.md}}.

\subsection{``Draft + Notes''}
This UI idea allows users to open multiple tabs, as in the previous one. However, this time, (at least) one of the tabs offers space for notes, that is, text that is not intended to be a part of the final text (i.e. non-diegetic text). For example, the user might write down a ``todo'' list or open questions related to their work on the draft. Again, this simple example idea does not involve AI.

\subsection{``Draft with AI feedback''}
This example introduces AI. Here, the user can write a draft in the left part of the UI, while the right part shows feedback by the AI. For instance, the AI might generate a comment on the text from the perspective of the intended target audience, as in the design by \citet{benharrak2023writerdefined}. Another example is the web app \textit{capito digital}, an ``AI tool for Easy Language''\footnote{\url{https://www.capito.eu/en/capitodigital}}. It gives feedback on a draft to make it easier to understand for people with less advanced skills in a particular language.

The double-headed arrow for this design idea in \cref{fig:ui_morphing} indicates that this involves both the \textit{Expose} and \textit{Inspire} pattern from \cref{fig:ui_patterns}F: Spelled out, the interpretation here is that the user's draft is exposed to the AI, which then gives feedback that may inspire the user to make changes to the draft.

\subsection{``Hover for local AI feedback''}
Similar to the previous example idea, the user receives feedback on their writing from the AI. However, here both text displays overlap: The main view is the writing area for the user. On top of that, a pop-up displays the AI's feedback. This pop-up appears when hovering over the part of the draft to which the feedback is related. %

\subsection{``Draft with AI suggestions''}
In this design idea, the user writes on a page while the AI shows diegetic text suggestions, that is, text snippets intended to be directly included in the draft. For example, the \textit{Wordcraft} system by \citet{yuan2022wordcraft} uses a such a UI. 

The double-headed arrow for this design idea in \cref{fig:ui_morphing} indicates that this involves both the \textit{Export} and \textit{Import} pattern from \cref{fig:ui_patterns}F: Interpreted for this example, we might say that the user's text is exported to the AI, so that it can provide fitting text continuations intended to then be imported into the draft by the user.

\subsection{``Prompt-to-draft''}
This UI idea focuses on AI writing, guided by the user. In particular, the user writes a prompt in the left text area, which triggers the AI to generate a fitting draft on the right. The user could then use this draft text elsewhere, beyond the depicted scope of this design. The single-headed arrow for this design idea in \cref{fig:ui_morphing} indicates that the prompt influences the generated text. That said, the user might of course also be influenced by reading that text (e.g. to then revise the prompt), which we could also annotate as an arrow in the opposite direction. For this example, we chose not to do so, to indicate that the main intended relationship in this design is to guide the system. 

A similar idea is used in \textit{Daramatron} by \citet{mirowski2023drama}; for example, their system generates whole dialogues/scenes from initial higher-level prompts. 

\subsection{``Paragraph-wise prompt-to-draft''}
This idea is very similar to the previous one. The key difference is that this UI offers a list of text areas, for example, to write one prompt per paragraph. Consequently, the AI generates a draft paragraph by paragraph. For example, a similar idea appears in the scene-wise prompting in one of the designs proposed by \citet{biermann2022tooltocompanion} (see therein: Figure 1c).

\subsection{``Digital post-it board''}
This example idea illustrates a simple case of using the ``spatial'' patterns from \cref{fig:ui_patterns}D and E: The user can enter text in multiple text areas, which can be arranged in 2D space on the screen. For example, this is possible in digital whiteboards, such as \textit{miro}\footnote{\url{https://miro.com}}.

\subsection{``Brainstorm-board with AI suggestions}
This design idea extends the previous one with AI suggestions. Concretely, the AI can add new post-it notes that extend the text/ideas written in the user's notes. This could make use of the spatial information, for example, such that the AI notes appear next to the user notes that served them as input. This direction of influence is indicated by the single-headed arrow for this design idea in \cref{fig:ui_morphing}.

\subsection{``Prompting board''}
Finally, this last design idea is very similar to the previous one but here the user is meant to enter non-diegetic text in their notes, that is, instructions to the AI (e.g. ``Suggest five video ideas for my HCI YouTube channel''). The AI responds by adding new notes, in the proximity of the user's corresponding instruction note. Again, this direction of influence is indicated by the single-headed arrow for this design idea in \cref{fig:ui_morphing}.

The web app \textit{Fermat}\footnote{\url{https://fermat.app}} uses (a family of) related UI and interaction concepts.

\section{Meta Reflection}\label{sec:appendix_meta}
\begin{figure*}
    \centering
    \includegraphics[width=.75\linewidth]{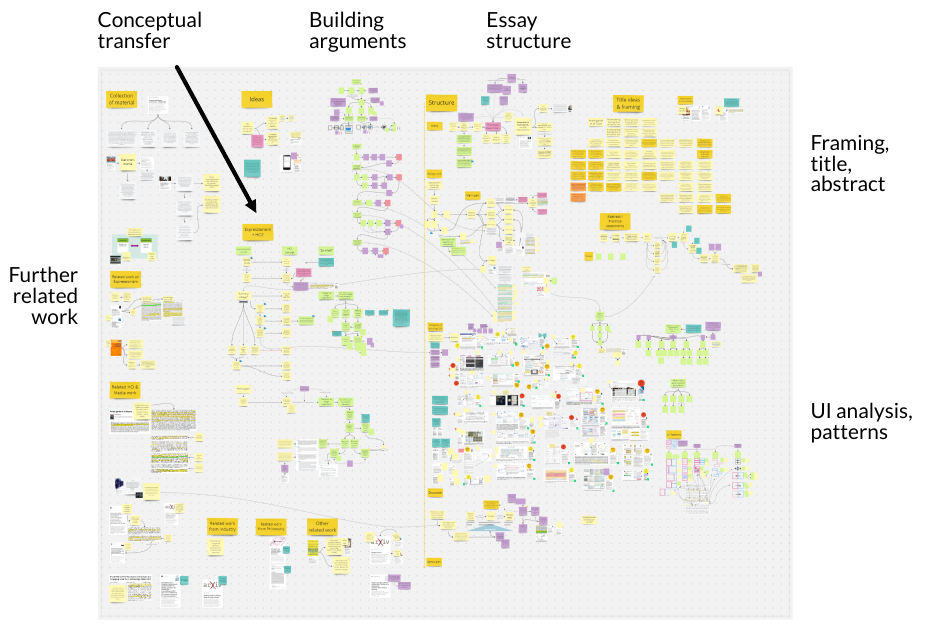}
    \caption{\revised{This essay was largely created on the shown \textit{miro board}, which extensively involved writing in fragments and using their spatial arrangement and arrows to externalise ideas and associations.}}
    \Description{Screenshot from a miro board from the software ``miro''. It is a canvas with many post-it notes, arrows, images, stickers, and so on. Annotations around the screenshot indicate which part of this board was used for what. In clockwise direction: ``Conceptual transfer'', ``Building arguments'', ''Essay structure'', ``Framing, title, abstract'', `UI analyis, patterns'', ``Further related work''.}
    \label{fig:meta}
\end{figure*}

\revised{As a ``meta illustration'' of the discussed concepts, we disclose the role of fragmentation in the development of this essay itself.} 
\revised{Concretely, the essay was largely conceptualised and created on a \textit{miro board}, shown in \cref{fig:meta}, which extensively involved writing in fragments and using their spatial arrangement and arrows to externalise ideas and associations. No AI tools were used for writing this essay, except for \textit{DeepL}\footnote{\url{https://www.deepl.com/}} to iterate on selected sentences for final polish.} 

\end{document}